\documentclass[12pt, draftclsnofoot, onecolumn]{IEEEtran}
\IEEEoverridecommandlockouts
\usepackage{hyperref}
\usepackage{threeparttable}
\usepackage{marvosym}
\usepackage{hyperref}
\usepackage{rotating}
\usepackage{multirow}
\usepackage{cite}
\usepackage{subfigure}
\usepackage{amsmath,amssymb,amsfonts}
\usepackage{algorithmic}
\usepackage{graphicx}
\usepackage{textcomp}
\usepackage{xcolor}
\usepackage{booktabs}
\usepackage{color}
\def\BibTeX{{\rm B\kern-.05em{\sc i\kern-.025em b}\kern-.08em
		T\kern-.1667em\lower.7ex\hbox{E}\kern-.125emX}}
\begin{document}
	
\title{Multi-task Learning-based CSI Feedback Design in Multiple Scenarios
	}

\author{
\normalsize {Xiangyi~Li,
Jiajia~Guo, 
Chao-Kai~Wen, \IEEEmembership{\normalsize {Senior Member,~IEEE}},\\
Shi~Jin, \IEEEmembership{\normalsize {Senior Member,~IEEE}},
Shuangfeng~Han, \IEEEmembership{\normalsize {Senior Member,~IEEE}}, and
Xiaoyun~Wang
}
\thanks{X.~Li, J.~Guo, and S.~Jin are with the National Mobile Communications Research Laboratory, Southeast University, Nanjing, 210096, P. R. China (email: Xiangyi\_li@seu.edu.cn, jiajiaguo@seu.edu.cn, jinshi@seu.edu.cn).}
\thanks{C.-K.~Wen is with the Institute of Communications Engineering, National Sun Yat-sen University, Kaohsiung 80424, Taiwan (e-mail: chaokai.wen@mail.nsysu.edu.tw).}
\thanks{S.~Han is with the China Mobile Research Institute, Beijing 100053, China (e-mail:
hanshuangfeng@chinamobile.com).}
\thanks{X.~Wang is with the China Mobile, Beijing 100032, China (e-mail: wangxiaoyun@chinamobile.com).}
}

\maketitle

\begin{abstract}

For frequency division duplex (FDD) systems, downlink channel state information (CSI) feedback is essential.  Deep learning-based auto-encoder (AE) structures have shown promise in reducing feedback overhead.  
However, designing a super-large AE network to handle the CSI of all scenarios is not practical.  
A more practical approach is to divide the CSI dataset by region/scenario and use multiple simple AE networks. 
However, this method requires high memory capacity, making it unsuitable for low-end user equipment (UE). 
In this paper, we propose a new UE-friendly framework based on multi-tasking mode. Our framework, called single-encoder-to-multiple-decoders (S-to-M), uses multi-task-learning to design multiple independent AEs into a joint architecture with a shared encoder that corresponds to multiple task-specific decoders.  
We also integrate GateNet as a classifier to enable the base station to autonomously select the right task-specific decoder for the subregion. 
Our experiments on a simulated multi-scenario CSI dataset show that our proposed S-to-M framework outperforms other benchmark modes by significantly reducing model complexity and UE memory consumption.
\end{abstract}


\IEEEpeerreviewmaketitle

\section{Introduction}
\label{introduction}

\subsection{Background}
Massive multiple-input and multiple-output (MIMO) is considered one of the core technologies in the fifth-generation (5G) communication system and will continue to be important in the future 6G\cite{9170653}. It improves system capacity and spectral efficiency, ensuring the transmission characteristics of ``high dimension, high capacity, more dense network and lower delay'' of 5G communication system\cite{5595728,6798744}. In massive MIMO systems, accurate knowledge of the downlink channel state information (CSI) at the base station (BS) is essential to ensure high-quality precoding work and guarantee efficient transmission \cite{6798744}. 
However, as the number of antennas deployed at the BS in massive MIMO systems continues to grow, providing feedback becomes challenging due to the vast CSI matrix \cite{4641946}. To reduce the CSI feedback overhead, efficient compression of CSI at the user equipment (UE) is necessary before feeding it back and reconstructing it at the BS to recover the original high-dimensional CSI. 

The CSI features reflect the channel characteristics and can be considered a high-dimensional and low-rank image. Thus, the mission of CSI feedback is to solve the problem of high-dimensional and low-rank image compression and reconstruction. Traditional methods based on compressed sensing (CS) \cite{kuo2012compressive} are time-consuming and inefficient because they are iterative algorithms that struggle with high-dimensional nonlinear problems. Furthermore, these CS-based methods assume strict assumptions on channels, such as channels being sparse in some bases, which may not hold in real systems, making it challenging to deploy these methods in practice \cite{2018CsiNet}.

The deep learning (DL)-based channel estimation methods \cite{dong2019deep, ma2020data} and precoding methods \cite{sohrabi2020robust, gao2022data} treat the CSI matrices as images and apply DL for feature extraction, which is consistent with CSI feedback \cite{jiaOverview}.
 Most DL-based CSI feedback works utilize an end-to-end based autoencoder (AE)\cite{zhai2018autoencoder} architecture, including an encoder network at the UE for the  dimensional reduction of the high-dimensional CSI matrix that outputs the compressed code before feedbacking to the BS, and a decoder network to reconstruct the original CSI from the compressed code. The AE neural network (NN) is trained offline via massive CSI samples and then deployed in the real system matching the training data. CSI compression, feedback and reconstruction involve complicated, high-dimensional, nonlinear and non-convex problems. Unlike the traditional CS-based methods with slow convergence rate, the DL-based data-driven methods can greatly improve reconstruction accuracy and feedback efficiency via their excellent fitting ability toward data representation. 


\subsection{Related work}

\par The DL-based CSI feedback framework was first introduced by CsiNet\cite{2018CsiNet}, which demonstrated the significant superiority of NNs over traditional CS-based methods. Following CsiNet, many DL-based works emerged to explore the potential of DL, focusing on several aspects, including: 
\begin{itemize}
    \item \textbf{NN architecture design.} CRNet\cite{lu2020multi} was designed with a multi-resolution structure, while \cite{2020GAN} adopted a deep convolutional generative adversarial network (DCGAN) that performs well with an extremely low compression ratio. \cite{wang2022transformer} benefits from the transformer architecture and achieves great performance.
    \item \textbf{Quantization module optimization.} CsiNet+\cite{guo2020convolutional} designed the quantization module to better apply in practice and a special offset network to compensate for quantization distortion. EfficientFi\cite{2021EfficientFi-Yang} adopted a vector quantized variational autoencoder (VQ-VAE) to design a trainable discrete codebook for quantization.
    \item \textbf{Correlation exploring.} CsiNet-LSTM \cite{2018Time-varying} extracted the time correlation in the time-varying channel via the recurrent NN structure. CAnet \cite{2022Uplink-aided} designed an uplink-aided CSI acquisition framework utilizing the correlation between downlink and uplink CSI. Distributed DeepCMC \cite{mashhadi2020distributed} benefits from the correlations among the channel matrices of nearby users to further improve performance.
    \item \textbf{Multi-module joint design.} \cite{sohrabi2021deep} proposed an end-to-end feedback framework, including distributed channel estimation, feedback, and downlink multiuser precoding. \cite{ma2021model} jointly designs the channel estimation and CSI feedback via a model-driven approach.
\end{itemize} 
Most of the mentioned works use the dataset in \cite{2018CsiNet}, i.e., numerical simulation dataset in indoor and outdoor scenarios generated by the COST2100 channel model\cite{COST2100}. Some competitions of wireless communication combined with artificial intelligence \cite{xiao2021ai} verify the application of neural network model in actual communication system by using measured CSI data.

\subsection{Problem definition}
In reality, the works mentioned above are all focused on improving model performance, and the models are specifically designed for a particular dataset. 
Due to the data-driven nature of DL, the performance of the NNs can be directly influenced by the CSI dataset, including sample number, sample similarity, and feature complexity or sparsity. These factors are determined by the scenario's complexity or sampling range/density. 
For instance, the same model, CsiNet\cite{2018CsiNet} has different performances on the COST2100 indoor and outdoor scenario datasets because of the varying scenario complexities and sampling range. However, there is currently no uniform standard for CSI dataset collection, and research on how these factors affect NN performance is missing. As a result, guidance and suggestions for practical deployment are lacking, particularly in the face of manifold complex scenarios. 
\begin{table}[t]
	\centering
	\setlength{\abovecaptionskip}{0cm}
    \setlength{\belowcaptionskip}{-0.2cm}
	\caption{\label{performance-range} NN's\tnote{2} performance VS CSI sampling range\tnote{1}.} 

\begin{threeparttable}
	\begin{tabular}{c|cccc}
		\hline \hline
		CSI Sampling range &  1m      & 10m    & 40m   & 200m\\
		NN's performance &  -22.34dB       & -5.08dB &-1.90dB &-0.77dB\\\hline \hline       

	\end{tabular}
 \begin{tablenotes}
        \footnotesize
        \item[*] The NN tested is CsiNet\cite{2018CsiNet} with compression ratio of $1/256$.
        \item[*] NN's performance is evaluated using the normalized mean square error (NMSE) measured in dB.
 \end{tablenotes}
\end{threeparttable}

\vspace{-0.8cm} 
	
\end{table}

In this work, we aim to bridge the gap between DL-based methods and practical deployment by exploring DL's potential from the perspective of the CSI dataset, rather than solely focusing on AE architecture design. We have observed a negative correlation between NN performance and the CSI sampling range. The results on a simulation dataset\footnote{
Each dataset in our study consists of 50,000 CSI samples. The positions of the UE are randomly sampled within a circular area centered 100 meters east of the BS. We vary the sampling range by changing the radius of the circle from 1 m to 200 m for comparison.} generated by QuaDRiGa\cite{QuaDRiGa} software are shown in Tab. \ref{performance-range}. As the sampling range increases and includes more scenarios, it increases the difference between the CSI samples, which results in poor network performance. There are two solutions to deal with the poor performance: increasing network capacity (complexity or compression ratio) and narrowing the CSI sampling range.

When deploying in a regular urban micro-cell with various scenarios, a NN of general complexity may struggle to handle the CSI sampled throughout the entire cell with all of its scenarios. However, the NN may perform well in a local subregion with only one or a few scenarios. Thus, the challenging tasks of CSI compression and reconstruction in the whole global cell can be divided into several manageable sub-tasks by splitting the dataset according to the sampling region.
Based on this observation, there are two modes of deployment\footnote{In this study, we focus on the case where low-end UE devices are used. All the NNs are trained offline before deployment, and there is no online adaptation or fine-tuning during deployment. This process is different from the online-adjustment methods described in \cite{TransferCSI2021Zeng}.}: (1) Designing a more complicated NN to handle the global cell's CSI. (2) Dividing the global cell into several local sub-areas and using multiple relatively simple NNs, with each responsible for a single local area. However, both solutions require high computing power or memory requirements for UE, making them unsuitable for low-level equipment such as IoT sensors. For mode 1, a complex AE network contains a complex encoder, which requires high computing power for UE. For mode 2, UE needs a lot of memory to store multiple sets of encoder parameters and switch to the corresponding encoder network when it moves to another subregion.

\subsection{Our contributions}
In our work, we present a new deployment mode that is designed to be more friendly to low-end UE devices and address the limitations of modes 1 and 2. When handling CSI from multiple scenarios, the feedback work of all scenarios CSI can be divided into multiple sub-tasks that are easy to handle according to the way of regional division.  We introduce a multi-tasking mode that involves dividing the CSI dataset and feedback by region and propose the single-encoder-to-multiple-decoders (S-to-M) mode. In this mode, a general encoder is used for CSI compression, which corresponds to multiple task-specific decoders for CSI reconstruction, based on multi-task-learning (MTL). The shared encoder can provide excellent performance and generalization ability with low complexity, making it convenient for the UE by requiring minimal memory and providing relatively stable network performance. We also introduce a GateNet at the BS, which is trained as a classifier to select the appropriate decoder.
The  major contributions of this work can be summarized as follows:
\begin{itemize}
    \item \textbf{CSI dataset segmentation}: 
     We demonstrate the feasibility of dividing the CSI dataset into multiple subregions based on the sampling areas. We analyze the distribution of CSI features (1-D and 2-D) and the sample frequency and correlation to measure task correlation, which lays the foundation for the multi-tasking mode.
     
    \item \textbf{MTL application}: 
    To reduce UE's computing power and storage capacity consumption, we propose a new UE-friendly mode, called S-to-M, based on MTL. We use the parameter-sharing mechanism to take advantage of task correlation, allowing a shared encoder to correspond to multiple task-specific decoders. Our MTL-based S-to-M mode can significantly reduce the NN's complexity and memory consumption in UE, while maintaining high performance. 

    \item \textbf{Autonomous switching decoder NN at BS}: Although our S-to-M mode uses a general encoder to solve the encoder NN switching problem when the scenario changes, the BS still faces the challenge of selecting the correct task-specific decoder NN. To address this, we complete our S-to-M mode with GateNet, a classifier at the BS that can identify the task number from the encoder's output and switch the decoder NN autonomously. The results show that the GateNet's supervised-learning performance of a simple NN is over 99.5\%, resulting in negligible performance loss to the overall framework.

\end{itemize}


Parallel to our work, Zhang et al. \cite{Zhang2023MTL} also tackled the multi-scenario CSI compression problem using MTL. However, we differ in our training approach. They pretrain S-to-S's parameters and fine-tune the decoder's parameters, while we train S-to-M end-to-end and jointly train the shared encoder and task-specific decoder without fixing encoder parameters. Our work presents the advantage of a parameter hard sharing mechanism, which provides strong generalization, and an interpretability analysis in Section \ref{Interpretability}.

\par The rest of this paper is organized as follows: Section \ref{system model} introduces the system model and the CSI feature analysis.
Section \ref{MTL-feedback} presents the
proposed MTL-based S-to-M mode for CSI feedback deployment, including the mechanism of MTL, the complete framework and complexity analysis.
Section \ref{Simulation-Results} provides CSI simulation details and the evaluation of our proposed S-to-M compared with the benchmarks.
Section \ref{conclusion} gives the concluding remarks.
\par \emph{Notations:} Vectors and matrices are denoted by boldface lower and upper case letters, respectively. $(\cdot)^{\rm{H}}$ and $\sf{cov}(\cdot)$ denote Hermitian transpose and the covariance, respectively. $\mathbb{C}^{m\times n}$ or $\mathbb{R}^{m\times n}$ denotes the space of $m\times n$ complex-valued or real-valued matrix. For a 2-D matrix $\mathbf{A}$, $\mathbf{A}[i,j]$, $\sf{row}_k(\mathbf{A})$ and $\sf{col}_k(\mathbf{A})$ represent the $(i,j)^{th}$ element, the $k^{th}$ row and $k^{th}$ column in matrix $\mathbf{A}$. For a 1-D vector $\mathbf{a}$, $\mathbf{a}[i]$ denotes its $i^{th}$ element. $||\cdot||_2$ is the Euclidean/L2 norm. $[a_k]_{k=1}^N$ represents a list of $[a_1,a_2,...,a_N]$. $\sf{para}(\cdot)$ and $\sf{FLOPs}(\cdot)$ denote the NN's parameter amount and the floating-point operations (FLOPs) amount\footnote{An addition and a multiplication count as a FLOPs operation in this paper.}.


\section{System Model}
\label{system model}
\subsection{Massive MIMO-OFDM FDD system}
\label{system model1}
Consider a typical urban micro-cell in an frequency-division duplexing (FDD) massive MIMO system with one BS serving for multiple single-antenna UEs. The BS is localed at the center of the cell and equipped with an $N_{\rm t}$-antenna uniform linear array (ULA)\footnote{We adopt the ULA model here for simpler illustration, While the analysis and the proposed model are not restrict to any specific array shape.}. We utilize orthogonal frequency division multiplexing (OFDM) in downlink transmission over $N_{\rm f}$ subcarriers. The received signal on the $n^{th}$ subcarrier for a UE can then be modeled as \cite{2018CsiNet}:
\begin{equation}
y_n = {\bf h}_n^{H}{\bf v}_{n}x_n  + z_{n},
\label{eq1}
\end{equation}
where ${\bf h}_n \in \mathbb{C}^{N_{\rm t}}$,  $x_{n} \in \mathbb{C}$ and  $z_{n} \in \mathbb{C}$ denote the downlink instantaneous channel vector in the frequency domain, the transmit data symbol and the additive noise, respectively. The beamforming or precoding vector $ {\bf v}_n \in \mathbb{C}^{N_{\rm t}}$
should be designed by the BS based on the received downlink CSI\cite{2021Beamforming}. We stack all the $N_{\rm f}$ frequency channel vectors and derive the downlink CSI metrix ${\bf H}_{S-F}=[{\bf h}_1,{\bf h}_2,...,{\bf h}_{N_{\rm f}}]\in \mathbb{C}^{N_{\rm t}\times N_{\rm f}}$ in the spatial-frequency domain.
\par We perform the same data preprocessing as in \cite{2018CsiNet}: 2D discrete Fourier transformation to obtain the sparse CSI in the angle-delay domain. The large dimensional CSI matrix can be sparse in the angle-delay domain due to the limited multipath time delay and the sufficient antennas at BS\cite{2018CsiNet}. Subsequently, we retain the first $N_{\rm c}$ ($N_{\rm c}<N_{\rm f}$) non-zero rows to further reduce the feedback overload and derive the dimension-reduced CSI in the angle-delay domain: ${\bf H}_{A-D} \in \mathbb{C}^{N_{\rm t}\times N_{\rm c}}$. Notice that in this paper, all the grayscale images of the angle-delay CSI refers to  ${\bf H}_{A-D}$. The complex-valued elements in ${\bf H}_{A-D}$ are also divided into real and imaginary real-valued parts, then normalized in the range of [0,1], and finally, we obtained the NN's input: ${\bf H}\in \mathbb{R}^{N_{\rm t}\times N_{\rm c}\times 2}$.

\subsection{Feature description of CSI in multiple scenarios}
\label{section:CSI feature analysis}

A typical urban micro-cell usually contains various scenarios. Each scenario has unique environmental characteristics, e.g., the distribution (density or height) of buildings, trees, and so on, depicting UE's surrounding scattering environment. Similar to \cite{2020Deep-Transfer}, we divide the whole cell into several subregions ($T$ subregions). In each subregion, users are randomly distributed and share the same (or similar) propagation environment. The BS is located at an elevated position in the cell with few surrounding scatterers \cite{2018Channel-Estimation}. Thus, the channel between BS and UE is much more affected by UE's surrounding scatterers than BS's, as depicted in Fig. \ref{fig:two-scenarios}. 

\par For the $u^{th}$ UE, we can obtain the multi-path frequency response channel vector ${\bf h}_n$ on the $n^{th}$ subcarrier in (\ref{eq1}) according to the clustered response model\footnote{The CSI dataset for all the experiments in this paper is simulated by  QuaDRiGa\cite{QuaDRiGa} and COST2100\cite{COST2100}, which are both part of the geometry-based stochastic channel models (GSCMs) that are described by the clustered response model.}:       
\begin{figure}
	\centering
 	\setlength{\abovecaptionskip}{0cm}
    \setlength{\belowcaptionskip}{-0.2cm}
	\subfigure[Propagation layout between different scenarios.\label{fig:two-scenarios}]{\includegraphics[height=2.7in]{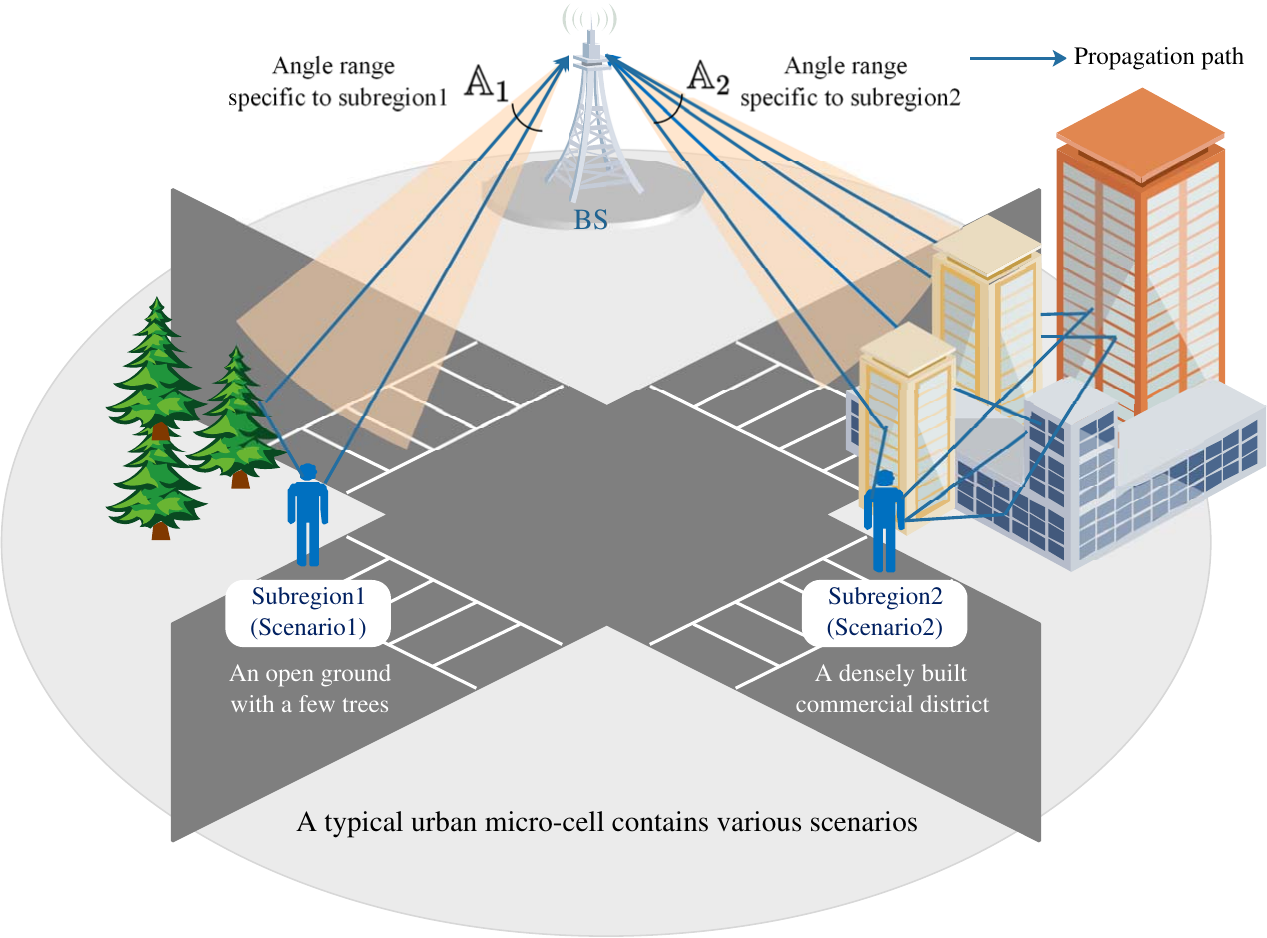}}
	\subfigure[Grayscale map of the corresponding angle-delay CSI.\label{fig:two-scenarios-graydraw}]{\includegraphics[height=2.7in]{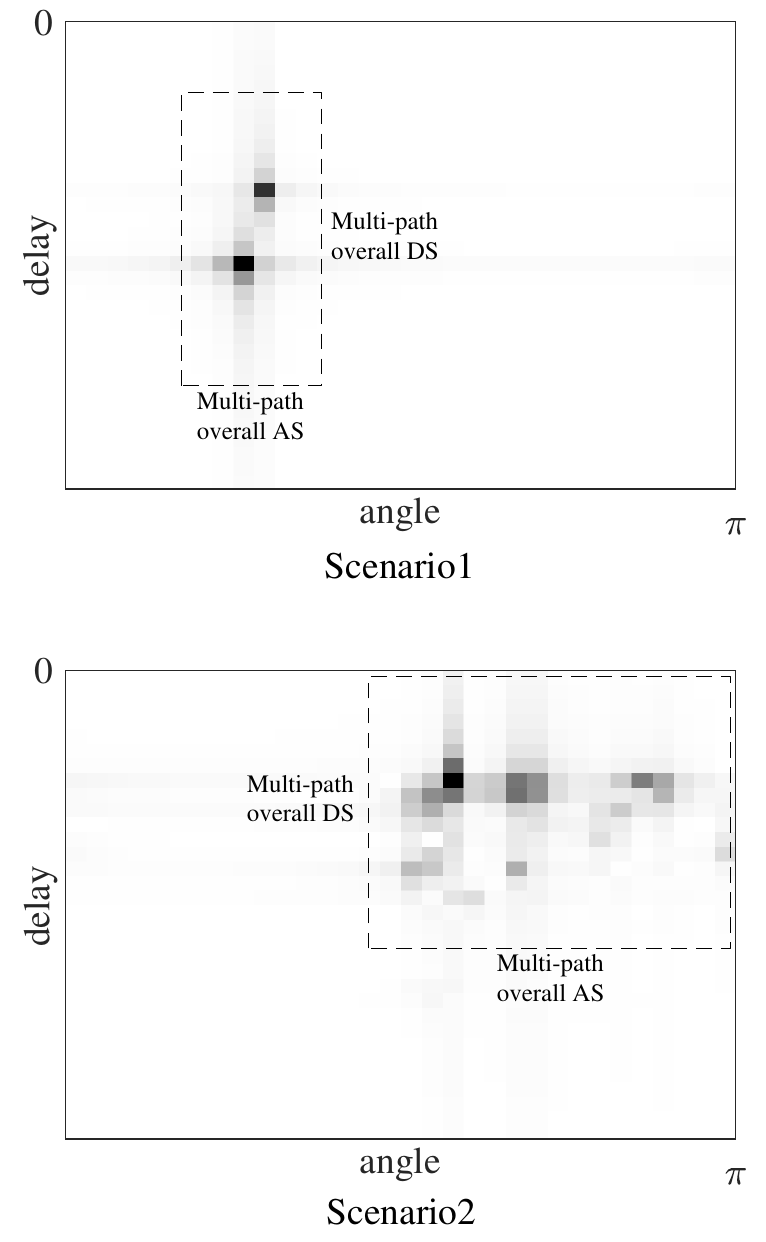}}
	\caption{A typical urban cell includes various scenarios (a), where the differences of the propagation layout can be reflected in the angle-delay CSI (b).}
	\label{fig:Path}
    \vspace{-0.8cm} 
\end{figure}

\vspace{-0.8cm} 
\begin{eqnarray}
    \mathbf{h}_{n} =\sum_{\ell=1}^{L} \,\, \iint\limits_{\tau_{\ell}\in \mathcal{T}_{\ell}, \phi_{\ell}\in \mathcal{A}_{\ell}} \alpha_{\ell} \mathrm{e}^{j(\theta_{\ell}-2\pi f_{n}\tau_{\ell})} \mathbf{a}_{\rm t}^{\rm H}\left(\phi_{\ell}\right)d\phi{_\ell}d\tau{_\ell},
    \label{eq2}
    \vspace{-0.8cm} 
\end{eqnarray}
where $L$ and $f_{n}$ denote the propagation path number and the $n^{th}$ subcarrier frequency with $\alpha_{\ell}$, $\theta_{\ell}$, $\tau_{\ell}$, $\phi_{\ell}$, $\mathcal{T}_{\ell}$ and $\mathcal{A}_{\ell}$ stand for the attenuation amplitude, random phase shift, delay, azimuth angles of departure (AoDs), delay's range, and AoD's range associated with the $\ell^{th}$ path, respectively. $\mathbf{a}_{\rm t}\left(\phi_{\ell}\right)$ stands for the antenna array response vectors at the BS, and the ULA antenna array response vectors can be given as \cite{2020Deep-Transfer}\cite{2018Channel-Estimation}\cite{2000system-model}:
\begin{eqnarray}
    \mathbf{a}_{\rm t}\left(\phi_{\ell}\right) =\frac{1}{N_{\rm t}} \left[1, \mathrm{e}^{-j \varpi_{n} \sin (\phi_{\ell})},..., \mathrm{e}^{-j \varpi_{n}({N_{\rm t}}-1) \sin (\phi_{\ell})}\right]^{\rm T},
    \label{eq3}
    \vspace{-0.8cm} 
\end{eqnarray}
in which $\varpi_{n}=2\pi d f_{n}/c$
with $c$ and $d$ denoting the speed of light and the distance between antenna elements, respectively.

\par The channel fading effects, such as the multipath or the spatial scattering, are captured in the CSI matrix, and the multipath propagation can be resolved in the angle-delay domain. 
In the clustered channel model, channels are expected to have limited $L$ scattering. Each cluster contributes a single main propagation path (represented by several sub-paths) between the BS and UE\cite{2013Spatially}\cite{2017Channel}, resulting in a small angular spread (AS) or delay spread (DS) of each path in the antenna or bandwidth resolution space\cite{QuaDRiGa}.

\par For the $u^{th}$ UE's $\ell^{th}$ propagation path, the AoD's range is given as $\mathcal{A}_{\ell}\triangleq [\bar{\phi}_{\ell}-\triangle_{\phi,\ell}, \bar{\phi}_{\ell}+\triangle_{\phi,\ell} ]$ with $\bar{\phi}_{\ell}$ and $\triangle_{\phi,\ell}$ denoting the mean $\phi_\ell$ and the AS of the $\ell^{th}$ path, as well as the delay's range: $\mathcal{T} _{\ell}\triangleq [\bar{\tau}_{\ell}-\triangle_{\tau,\ell}, \bar{\tau}_{\ell}+\triangle_{\tau,\ell}]$ with $\bar{\tau}_{\ell}$ and $\triangle_{\tau,\ell}$ denoting the mean $\tau_\ell$ and the DS. This can be observed in Fig. \ref{fig:Path}, where Fig. \ref{fig:two-scenarios} depicts two different scenarios and the propagation layout, and Fig. \ref{fig:two-scenarios-graydraw} shows the corresponding CSI's grayscale map simulated by QuaDRiGa. Especially for UE in scenario1 (an open ground), the propagation layout only contains a line of sight (LOS) path and a non-LOS (NLOS) path, reflected in the CSI map with the positions of the feature spots manifesting the different path lengths and AoDs between LOS and NLOS. The range of spots represent AS and DS\cite{2000system-model}. 
Moreover, for the $u^{th}$ UE, the overall multi-path angle range and delay range in Fig \ref{fig:two-scenarios-graydraw} can be expressed as: $\mathcal{A}_{u}\triangleq \bigcup_{\ell=1}^{L} \mathcal{A}_{u,\ell}$ and $\mathcal{T}_{u}\triangleq \bigcup_{\ell=1}^{L} \mathcal{T}_{u,\ell}$, respectively, with $\mathcal{A}_{u,\ell}$ and $\mathcal{T}_{u,\ell}$ specific to the $\ell^{th}$ path of the $u^{th}$ UE. As depicted in Fig \ref{fig:two-scenarios}, $\mathcal{A}_{u}$ and $\mathcal{T}_{u}$ can be covered by (or limited to) an angle range and a delay range upper bound  specific to the subregion where the UE is:
\begin{eqnarray}
    \mathbb{A}_k\triangleq\bigcup_{u\in U_k}\mathcal{A}_{u},\quad
    \mathbb{T}_k\triangleq\bigcup_{u\in U_k}\mathcal{T}_{u},
\label{Eq:range}
\end{eqnarray}
where $U_k$ denotes the index set of UE within the $k^{th}$ subregion. 

\begin{figure}[t] 
	\centering
	\setlength{\abovecaptionskip}{-0.2cm}
	\includegraphics[width=0.96\linewidth]{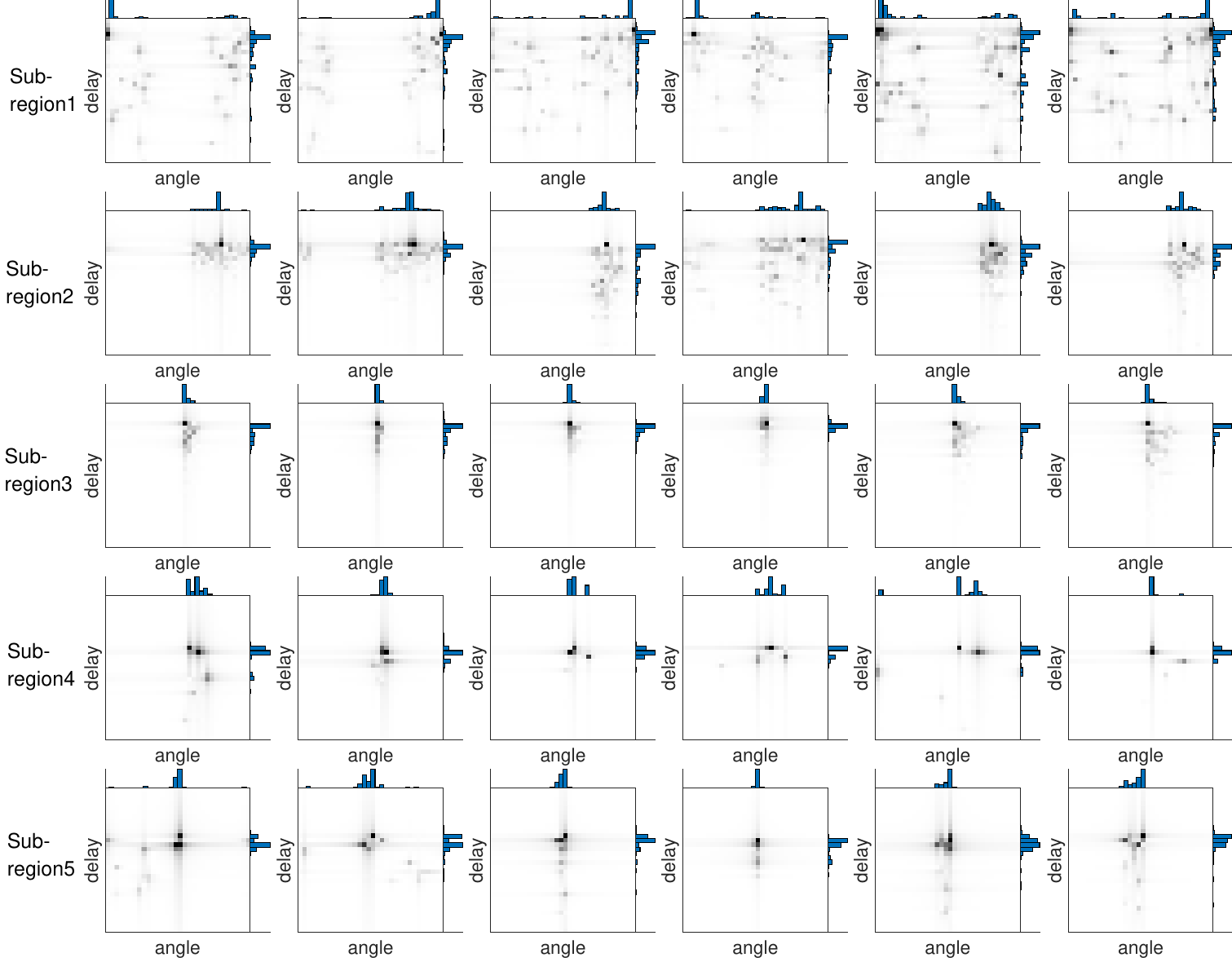}
	\caption{Grayscale image, PAS distribution, and PDP distribution of CSI maps in 5 subregions (subtasks). }
	\label{fig:5P-graydraw}
\vspace{-0.8cm} 
\end{figure}

This statistical distribution phenomenon can also be viewed in Fig. \ref{fig:5P-graydraw}, where the multi-scenarios CSI samples in five subregions are simulated by QuaDRiGa \cite{QuaDRiGa} with more details in Subsection \ref{subsubsec:Qua-dataset}. Each row contains six CSI samples randomly sampled in the corresponding subregion. In each CSI map, the power delay profile (PDP) and power angular spectrum (PAS) are depicted on the right and above the CSI image, representing the 1-D feature distributions, which can be calculated according to (\ref{eq2}-\ref{eq3}) as:
\vspace{-0.4cm} 
\begin{eqnarray}
\rm{PAS}(\mathbf{H})&=& \frac{1}{N_{\rm t}} \left [ \left \| \sf{col}_k(\mathbf{H} ) \right \|_2^2  \right ]_{k=1}^{N_{\rm t}}, \nonumber\\
\rm{PDP}(\mathbf{H})&=& \frac{1}{N_{\rm c}}\left [ \left \| \sf{row}_k(\mathbf{H} ) \right \|_2^2  \right ]_{k=1}^{N_{\rm c}},
\label{equ-PAS-PDP}
\end{eqnarray}
where $\mathbf{H}$, $||\cdot||_2$, $\sf{col}_k(\cdot)$ and $\sf{row}_k(\cdot)$ denote the CSI matrix, the Euclidean norm, and $k^{th}$ column and row of matrix, respectively. 
Within each row (subregion), the 1-D and 2-D CSI features are distributed over a specific large or small range, determined by the spatial environmental characteristics of the corresponding subregion. For example, the shape of the scattering areas and spatial density of the scatterers will impact the range of the CSI features. A rich scattering environment, such as a commercial district (subarea 1 in Fig. \ref{fig:5P-graydraw}), is reflected in dense and complex CSI with a wide angular and delay range. On the other hand, the CSI map of an open scattering environment, like a park (subarea 3 in Fig. \ref{fig:5P-graydraw}), will be sparse, with the feature centrally distributed.

\subsection{Subtask correlation analysis}
\label{Subtask correlation}
\begin{figure}
	\centering
	\setlength{\abovecaptionskip}{-0.2cm}
	\includegraphics[width=1.0\linewidth]{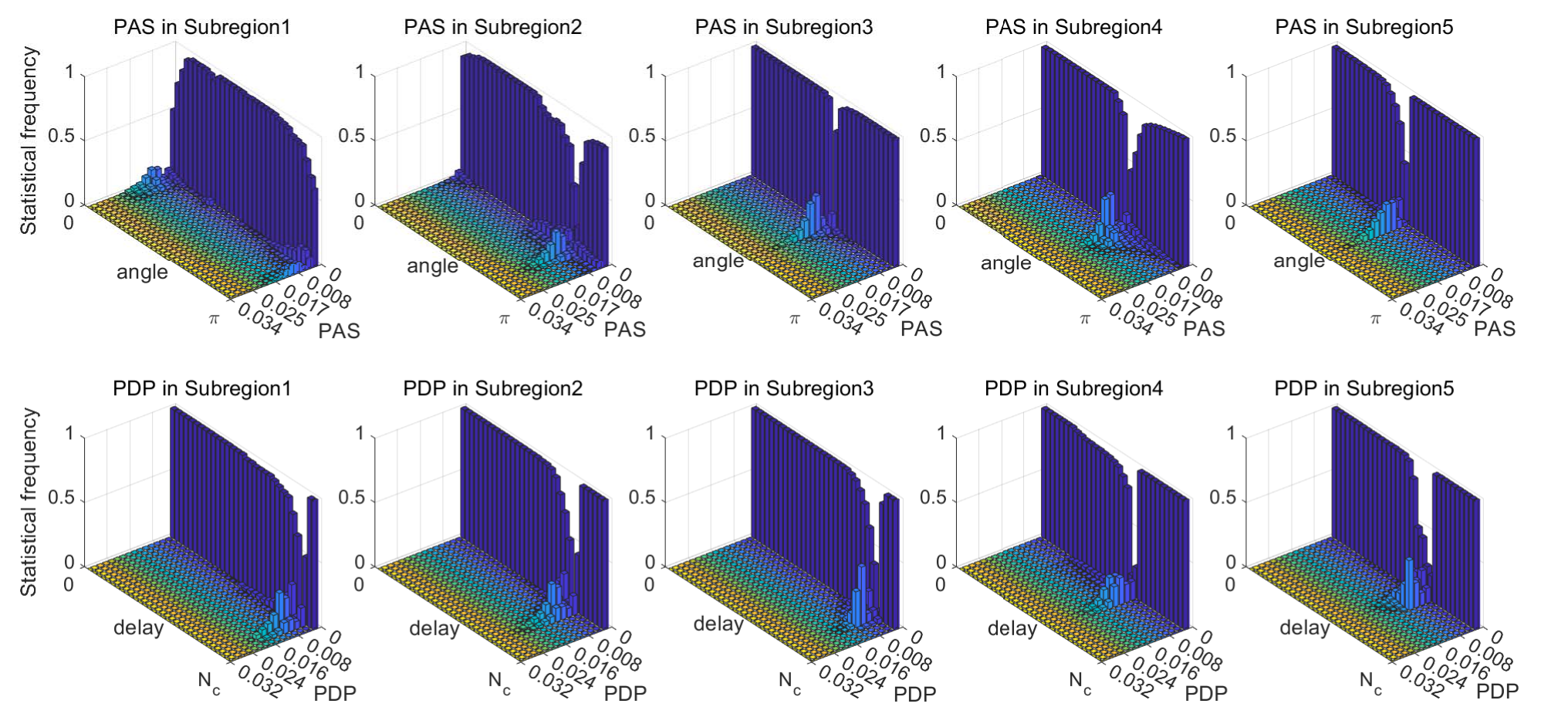}
	\caption{Statistical frequency distribution histogram of the 1-D feature (PAS and PDP) for CSI samples. }
	\label{fig:sample_frequency_5P}
 \vspace{-0.4cm} 
\end{figure}
As described in Subsection \ref{section:CSI feature analysis}, the CSI features are closely related to the UE's surrounding scattering environment, which justifies splitting the dataset by sampling region and lays the foundation for the multitasking modes, i.e., multiple subtasks for the feedback of subregion CSI. These subtasks are isomorphic, differing only in the CSI dataset. In this subsection, we focus on exploring the correlation of subtasks by analyzing the statistical properties of the multi-scenario CSI dataset. For example, we examine the sample's statistical frequency and similarity, which serve as the basis for applying MTL in our S-to-M mode.

Instead of displaying several samples' visualizations in Fig. \ref{fig:5P-graydraw}, we present the statistical frequency distribution histogram of the 1-D feature for CSI in Fig. \ref{fig:sample_frequency_5P}. In the figure, PAS is in the first row and PDP in the second row, with each column representing one subregion. We conducted statistical frequency analysis based on 50,000 randomly sampled CSI samples in each local subregion, which is sufficient to represent the overall CSI feature distribution of the subregion. We denote $\mathbb{D}_k$ as the dataset of the subtask $\mathcal{S}_k$ for the $k^{th}$ subregion's CSI feedback. The sample statistical frequency distribution of $\mathbb{D}_k$ is presented by its probability density function (PDF), $p^{\rm A}_{k}(\rm{PAS})$ for PAS and $p^{\rm D}_{k}(\rm{PDP})$\footnote{The PDF of the 1-D features can be seen as the marginal PDF of the 2-D feature, where the combination of PAS and PDP presents the 2-D CSI matrix. Here, we display the marginal PDF for the sake of visualization, as the 2-D feature vector is too long (1024-length) to straightforwardly capture its distribution characteristic.} for PDP, which can be regarded as an approximate representation of the overall feature distribution. That is, if $\mathbf{H}\in \mathcal{D}_k$ then $\rm{PAS}(\mathbf{H})\sim p^{\rm A}_{k}(\rm{PAS})$ and $\rm{PDP}(\mathbf{H})\sim p^{\rm D}_{k}(\rm{PDP})$. Thus, we can denote the upper bound of angular range $\mathbb{A}_k$ or delay range $\mathbb{T}_k$ in \ref{Eq:range} as the confidence interval in angular or delay domain, respectively, with the 95\% confidence level, i.e., $\mathbb{A}_k$ and $\mathbb{T}_k$ can cover most CSI samples' $\phi$ and $\tau$ except for the 5\% outliers. By comparing $p^{\rm A}_{k}(\rm{PAS})$, $p^{\rm D}_{k}(\rm{PDP})$, $\mathbb{A}_k$ and $\mathbb{T}_k$ for different subtasks in Fig. \ref{fig:sample_frequency_5P}, the CSI feature distribution varies across different scenarios.

\begin{figure}
	\centering
	\setlength{\abovecaptionskip}{-0.2cm}
	\includegraphics[width=0.75\linewidth]{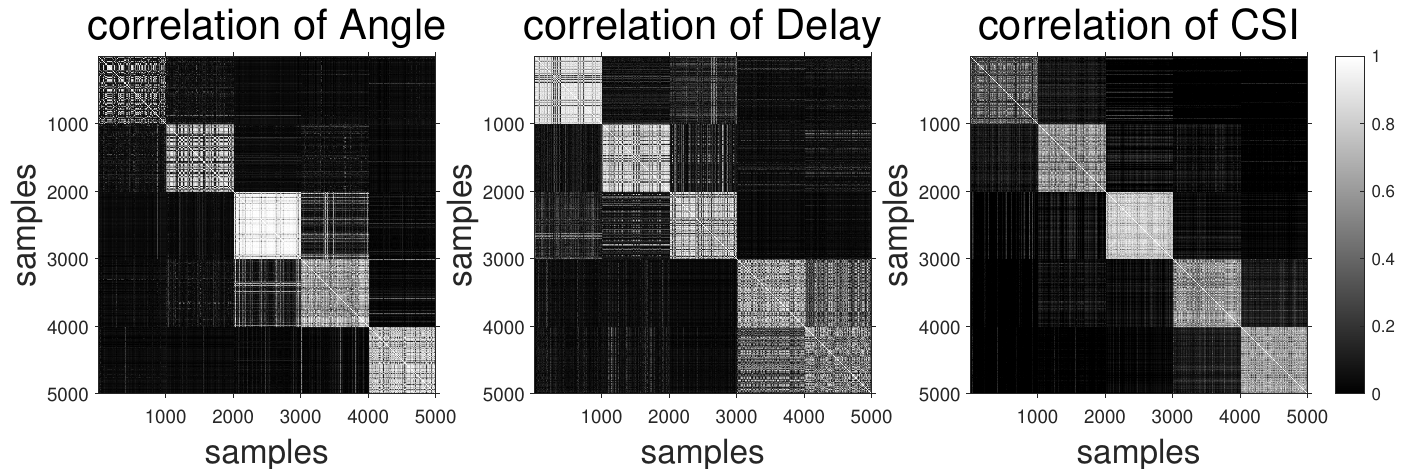}
	\caption{Left: correlation of PAS; middle: correlation of PDP; right: correlation of CSI matrix. }
	\label{fig:correlation_5P}
	\vspace{-0.8cm} 
\end{figure}
Furthermore, the CSI maps sampled within a local subregion exhibit higher correlations than those sampled across different local subareas. This finding explains why the performance of the subtask's neural network is better than that of the entire cell from the perspective of sample similarity. We introduce Pearson correlation to describe the similarity between CSI samples, where the $(i,j)^{th}$ element in the correlation matrix $\mathbf{R}$ can be expressed as:

\begin{eqnarray}
\mathbf{R}[i,j] = \frac{\sf{cov} (\mathbf{H}^{\mathit{i}}, \mathbf{H}^{\mathit{j}})}{\sqrt{\sf{cov}(\mathbf{H}^{\mathit{i}}, \mathbf{H}^{\mathit{i}}) \sf{cov}(\mathbf{H}^{\mathit{j}}, \mathbf{H}^{\mathit{j}})}} ,
\label{equ5}
\end{eqnarray}
where $\sf{cov}(\cdot)$ denotes the covariance and the superscript $i$ or $j$ stands for the sample number. Fig. \ref{fig:correlation_5P} shows the visualization of the sample correlation (Pearson correlation) matrix, which includes the PAS, PDP, and CSI matrix. The small square \textcolor{red}{matrices} on the diagonal are shallower than the rest, manifesting that the Pearson correlation of CSI samples within a subregion is higher than those crossing different regions. The sample correlation of CSI matrix can be viewed as the combination of PAS's and PDP's.

\par The following demonstrates the reasons for the above statistical phenomena of the CSI database: During the CSI simulation, we restrict the range of these subregions, represented by diameter, to less than the correlation distance (CSI sampled within this range will have a strong spatial correlation\cite{QuaDRiGa}.). As a result, the large-scale fading parameters (LSPs) within one local area do not change considerably and keep their wide-sense stationary (WSS) properties \cite{QuaDRiGa}. Therefore, the change of small-scale fading parameters (SSPs) is the main factor affecting the difference of CSI distribution in one subregion. Both the LSPs and SSPs change dramatically across different subregions, resulting in a much lower CSI correlation as well as the various $\mathbb{A}_k$ and $\mathbb{T}_k$. Thus, dividing the dataset according to the sampling area is reasonable and equivalent to an artificial sample clustering process.


\section{MTL-based CSI Feedback}
\label{MTL-feedback}

\begin{figure}
    \centering
    \subfigure[S-to-S\label{fig:s2s}]{\includegraphics[width=3.2in]{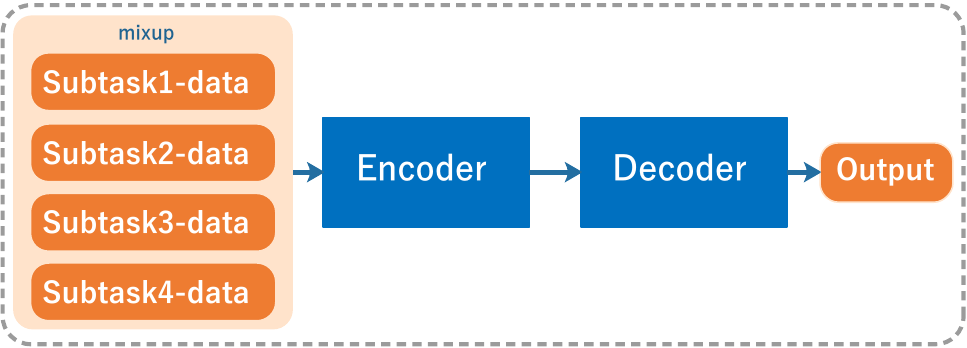}}
	\subfigure[M-to-M\label{fig:m2m}]{\includegraphics[width=3.2in]{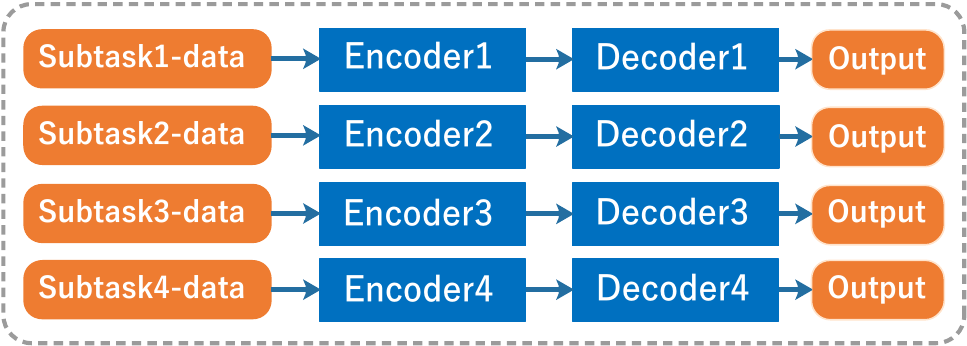}}
	\subfigure[S-to-M\label{fig:s2m}]{\includegraphics[width=4in]{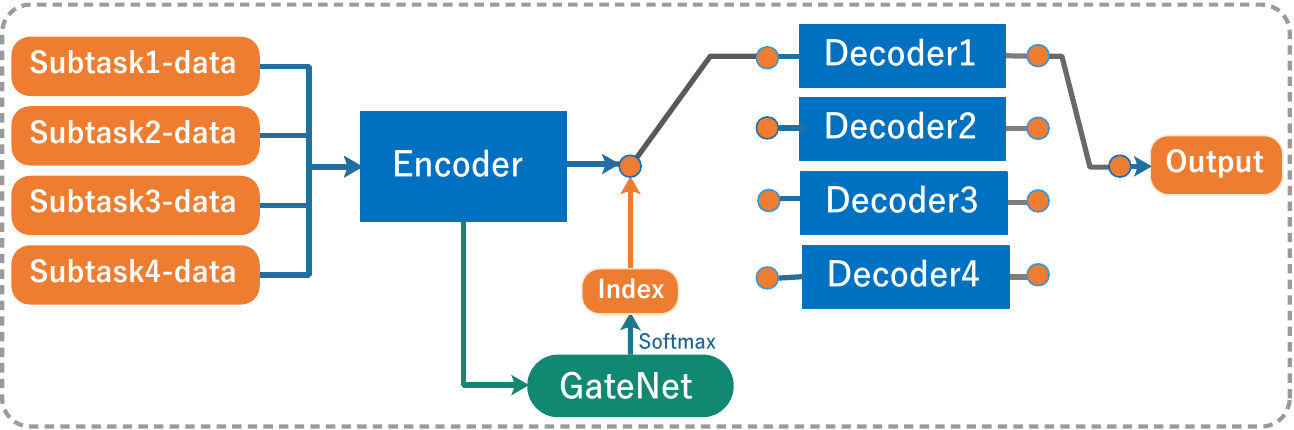}}
	\caption{S-to-S/M-to-M/S-to-M modes of deployment, where the number of sub-tasks is 4 as an example.}

    \label{fig:3-modes}
    
\vspace{-0.8cm} 
\end{figure}

In Section \ref{system model}, we analyze the CSI feature distribution and describe that decomposing a problematic task into multiple yet relatively easy sub-tasks is feasible by dividing CSI datasets according to sampling areas. 
Therefore, when dealing with CSI sampled from the whole cell with multiple scenarios, two deployment modes are usually adopted:
\begin{itemize}
    \item Single-encoder-to-single-decoder(S-to-S) mode (Fig. \ref{fig:s2s}): This mode involves designing a single sophisticated neural network with high complexity to handle the whole cell's CSI feedback.
    \item Multiple-encoders-to-multiple-decoders(M-to-M) mode (Fig. \ref{fig:m2m}): This mode involves dividing the CSI dataset and feedbacking it by region, using multiple relatively simple neural networks. Each neural network is responsible for the feedback of CSI sampled in the local sub-area.
\end{itemize}
Given the input $\mathbf{H}\in \mathcal{D}_k$, where $\mathcal{D}_k$ is the $k^{th}$ subtask's dataset, the output $\hat{\mathbf{H}}$ of the above two modes can be expressed as:
\vspace{-0.4cm}
\begin{align}
\hat{\mathbf{H}}_{\rm{S-to-S}} &= \sf{Dec}(\sf{Enc}(\mathbf{H};\Phi);\Psi), \nonumber\\
\hat{\mathbf{H}}_{\rm{M-to-M}} &= \sf{Dec}_k(\sf{Enc}_k(\mathbf{H};\Phi_k);\Psi_k), 
\label{equ7}
\end{align}
where $\sf{Enc}(\Phi)$ and $\sf{Dec}(\Psi)$ stand for the shared and common encoder and decoder, whereas $\sf{Enc}_k(\Phi_k)$ and $\sf{Dec}_k(\Psi_k)$ are correspond to the $k^{th}$ subtask. In M-to-M mode, the subtask index $k$ should be acquired both at the UE and BS, which involves the link of the additional index information feedback.

\par The S-to-S mode is unsuitable for users of low-end devices. For example, the hundreds of megabytes in the encoder network model designed in AI competition \cite{xiao2021ai} may not be deployed in low-end devices such as small sensors in practice. The M-to-M mode is also not friendly enough for low-end UE. It greatly reduces the complexity of NN, especially the encoder NN at UE, and saves substantial resource costs in NN architecture designs and parameter adjustment. However, it increases the link of switching network according to the environment and area in actual deployment. When the UE travels from one subregion to another, both the current encoder NN and decoder NN are unsuitable for the dramatically changed CSI owning to the limited generalization ability and must be altered to the corresponding NN. Moreover, the M-to-M mode requires UE to be equipped with ample memory to save multiple sets of encoder NN parameters, which burdens the UE significantly. Otherwise, the UE needs to download the new parameters frequently, which adds additional transmission burden and causes delays in the system. Furthermore, download errors can reduce the encoder NN's performance.

\subsection{Framework of S-to-M}
Our deployment mode comes from two aspects. On the one hand, although the M-to-M mode finds a balance between the model complexity and the CSI sampling range, it does not fully take advantage of MTL, because the multiple NNs are designed independently. In this case, a non-negligible task correlation exists between multiple tasks. The feature distribution of CSI sampled in different sub-areas will be affected by common factors, such as the number and distribution of BS antennas, the climate of the whole cell, the performance of transceiver equipment, and so on. Therefore, we introduce MTL to improve the M-to-M mode via jointly designing the multiple NNs to explore this task correlation and learn the sharing representations. On the other hand, both the S-to-S and M-to-M modes are not designed as UE-friendly. They either require high computational power or memory of UE or have a significant downloading burden.

Inspired by these two aspects, we propose a new MTL-based CSI feedback deployment mode, S-to-M, by utilizing the hard-sharing structure of MTL. In this mode, a common and shared encoder corresponds to multiple task-specific decoders, as shown in Fig. \ref{fig:s2m}. By exploring the task correlation through MTL, the shared encoder can achieve exceptional performance and generalization ability with low complexity.

\begin{figure}
    \centering
    \subfigure[Hard sharing parameter\label{fig:hard-sharing}]{\includegraphics[height=2in]{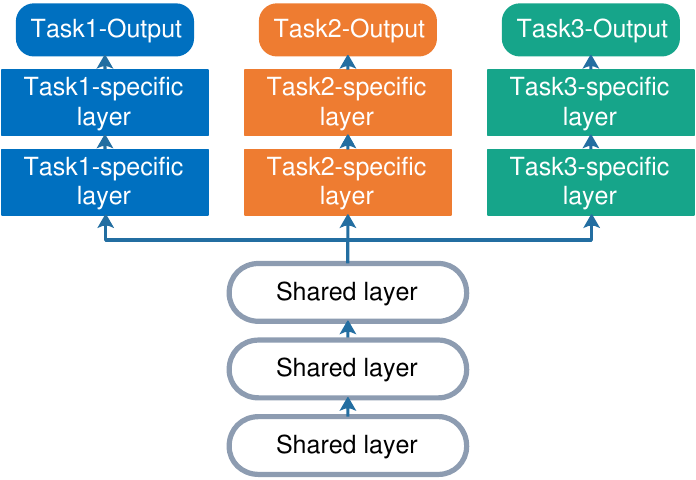}}
	\subfigure[Soft sharing parameter\label{fig:soft-sharing}]{\includegraphics[height=2in]{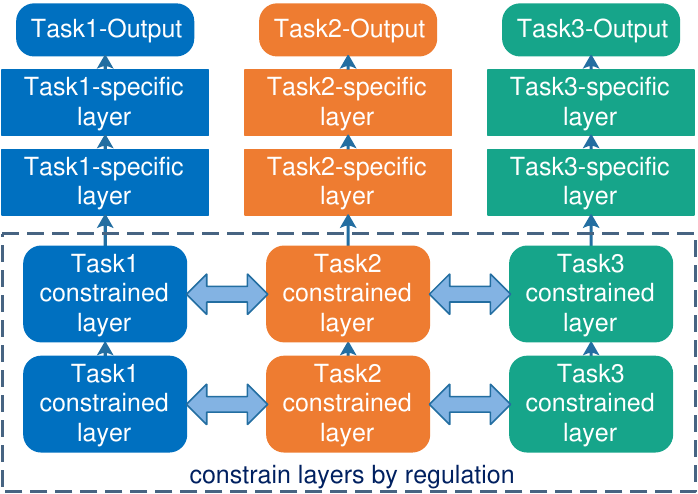}}

	\caption{Hard-sharing and soft-sharing mechanisms in MTL, with different tasks colored differently.}
    \label{fig:Hard-soft}
\vspace{-0.8cm} 
\end{figure}
\paragraph{MTL-based joint architecture design}

\par MTL is a relatively mature technology in machine learning and has made exemplary achievements in many fields. The basic idea of MTL is to combine multiple relative tasks to learn simultaneously to enhance each model's representation and generalization abilities, which is mainly realized by sharing representations between related tasks through the mechanism of sharing parameters \cite{ruder2017overview}. The primary parameters sharing mechanism of deep MTL are the two kinds depicted in Fig. \ref{fig:Hard-soft}: (1) Hard-sharing (Fig. \ref{fig:hard-sharing}) architecture stacks the task-specific layers on the top of the shallower shared layers \cite{collobert2008Hardsharing}; (2) Soft-sharing (Fig. \ref{fig:soft-sharing}) approaches allow each task has its parameters and model, where the distance between the part of the model's parameters is regularized to encourage parameter similarity \cite{misra2016cross}. Unlike the multiple independent sets of parameters in M-to-M, the regularized part of parameters in soft-sharing focus on learning the task correlation and sharing representations.

We have chosen the hard-sharing architecture to design our MTL-based CSI feedback mode as it is more suitable for closely related tasks and more appropriate for our UE-friendly-design needs. Moreover, the hard-parameter-sharing significantly reduces the overfitting risk of the shared parameters (specifically reduces to $1/T$ of the unshared parameters, where $T$ is the number of sub-tasks \cite{baxter1997bayesian}). The shared parameters can access the samples from all tasks. 
In contrast, the task-specific parameters can only access their subtask samples, making it challenging to find a representation that captures all tasks and making it equivalent to a kind of data enhancement. This assists the shared parameters in obtaining outstanding performance and generalization ability with low complexity without the risk of overfitting.

\par In our S-to-M mode, the shared encoder is designed as the sharing parameter part in the hard-sharing architecture to extract the sharing representation and exploit the task correlation, such as the factors of transceiver equipment's performance or the consistency of geography and climate in the whole cell. The multiple decoders stand for the top task-specific layers to learn the differences between the subtasks, e.g., LSPs and SSPs. 

\begin{table*}[t]
	\centering
	\setlength{\abovecaptionskip}{0cm}
    \setlength{\belowcaptionskip}{-0.2cm}
    \renewcommand\arraystretch{0.85}
	\caption{\label{Tab:GateNet} GateNet architecture.}
	\resizebox{0.38\textwidth}{!}{
    \begin{tabular}{ccc}
    \hline
    \multicolumn{3}{l}{\textbf{Input}: The feedback code $\mathbf{s}=Enc(\mathbf{H},\Phi)$} \\\hline
    \textbf{FC layers}          & \textbf{Input\_dim/Output\_dim}                         & \textbf{Activation}         \\
    1                 & dim($\mathbf{s}$)/2048          & BN+ReLU            \\
    2                 & 2048/512                                       & BN+ReLU            \\
    3                 & 512/$T$                                          & BN+Softmax         \\\hline
    \multicolumn{3}{l}{\textbf{Output}: the index one-hot label} \\\hline                                
    \end{tabular}}
\vspace{-0.8cm} 
\end{table*}
\paragraph{Autonomous switching decoder NN at BS}
The MTL-based CSI feedback mode involves the problem of switching the decoder networks. When the UE moves across different subregions, extra environmental information such as the UE's position should be fed back to the BS to guide its work, but this may be unavailable due to safety issues. To handle this problem, we improve our S-to-M deployment mode framework by enabling BS switching decoder NN autonomously. Specifically, we add a classifier NN, GateNet, at the BS to output the label from the encoder’s output and conduct BS to switch for the correct decoder networks. This way, no additional information is involved, which reduces the feedback overhead and simplifies the process of CSI feedback in multiple subregions.
\par The overall framework of S-to-M mode with GateNet is depicted in Fig. \ref{fig:s2m}, where the encoder at UE is equipped with one shared encoder NN and the decoder at BS is equipped with multiple task-specific decoder NNs (colored in blue). A simple classifier, GateNet, is also deployed at BS (colored in green) to identify the task category to which the CSI sample belongs through the compressed feedback code. The architecture of GateNet is shown in Tab. \ref{Tab:GateNet}, which consists of several fully connected (FC) layers. The last Softmax activity outputs the most likely category in the form of maximum probability. We also convert $g$ to the subtask index identified by GateNet via $g=\sf{argmax}(\mathbf{g})$, where $\mathbf{g}$ is the GateNet output and $g$ denotes the index number of the max-value element in $\mathbf{g}$. When deploying, the BS first uses GateNet to identify the subtask index once it receives the feedback code, and then selects the corresponding decoder for CSI decompression and reconstruction. Given the input $\mathbf{H}\in \mathcal{D}_k$, the output of the overall S-to-M mode with GateNet is expressed as:
\begin{align}
\hat{\mathbf{H}}_{\rm{S-to-M}}&= \sf{Dec}_{g}(\sf{Enc}(\mathbf{H};\Phi),\Psi_{g}), \nonumber\\
where\quad g &=\sf{argmax}(\sf{GateNet}\sf{Enc}(\mathbf{H};\Phi))).
\end{align}
Although the expression shows that the overall MTL-based CSI feedback framework is highly influenced by GateNet's classification accuracy, the experimental results demonstrate that the nearly perfect performance (accuracy over 99.9\%) indicates that the loss caused by GateNet is small enough to be disregarded. The encoder's output includes sufficient information for GateNet's classification task.

\subsection{Training}
\par
\begin{figure}[t] 
	\centering
	\setlength{\abovecaptionskip}{-0.2cm}
	\includegraphics[width=0.96\linewidth]{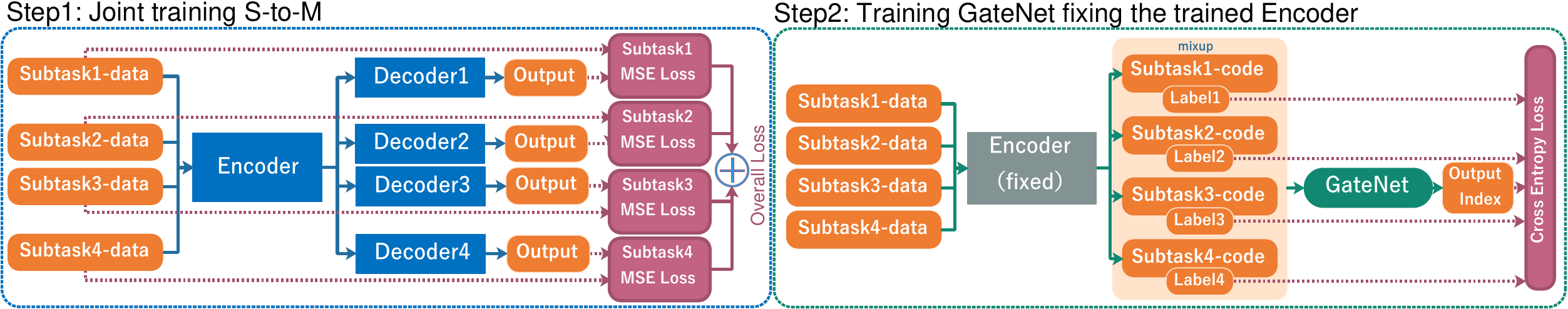}
	\caption{Two-step training of the overall framework of S-to-M mode.}
	\label{fig:training}
\vspace{-0.8cm} 
\end{figure}

 As depicted in Fig. \ref{fig:training}, the training of the overall framework can be divided into two steps: For the first step, we train the MTL-based joint architecture, i.e., jointly train the shared encoder and multiple decoders with the joint loss function:
\begin{eqnarray}
L_{\mathrm{MTL}} = \frac{1}{T}\sum_{k  =  1}^{T}\left (  \frac{1}{N_k}\sum_{n=1}^{N_k}\left \| \sf{Dec}_k(\sf{Enc}(\mathbf{H}_{k}^n;\Phi);\Psi_k)-\mathbf{H}_{k}^n \right \|_2^2 \right )  ,
\end{eqnarray}
where $T$ stands for the number of subtasks with the corresponding subscript $k$ and $N_k$ denotes the number of training samples for the $k^{th}$ subtask with the corresponding superscript $n$.
\par For the second step, we train the GateNet. We first fix the shared encoder's parameters $\Phi$ to obtain the encoder's output $\mathbf{s}_{k}^n=\sf{Enc}(\mathbf{H}_{k}^n,\Phi)$ and then label them with the corresponding task category with one-hot code $\mathbf{\ell}_k$, where the $k^{th}$ element in $\mathbf{\ell}_k$ is 1 whereas other elements are 0.
Subsequently, we derive the training data $\bigcup_{k=1}^{T} \left \{ (\mathbf{s}_{k}^n,\mathbf{\ell}_{k}) \right \} _{n = 1}^{N_{k}}$ of the classifier GateNet for supervised learning. A basic classification loss function, cross-entropy, is utilized to train GateNet. 
\begin{eqnarray}
\vspace{-0.2cm}
L_{\mathrm{GateNet}} = \sum_{k=1}^{T}\sum_{n=1}^{N_k}\left (  \sum_{i=1}^{T} \mathbf{\ell} _{k}[i]\log_2(\mathbf{g}_{k}^n[i]) \right ),
\end{eqnarray}
where $\mathbf{g}_{k}^n=\sf{GateNet}(\mathbf{s}_{k}^n)$ and $\mathbf{g}_{k}^n[i]$ denotes its $i^{th}$ element. 

\subsection{Complexity analysis}
Since the S-to-M mode can utilize NN with any architecture or complexity, we mainly discuss and compare the complexity between our proposed S-to-M mode and the S-to-S or M-to-M modes. For convenience, the multiple task-specific decoders with NN architectures are the same in the analysis, although the S-to-M mode is not restricted to this limitation. To be fair, the comparison is under the situation that the three modes obtain the same performance levels and feedback overhead, where the S-to-S mode applies a much more complicated encoder and decoder NN, denoted as $Enc_{\mathrm{com}}$ and $Dec_{\mathrm{com}}$ (combination denoted as $AE_{\mathrm{com}}$), whereas the encoder
and decoder NN in the mode with multiple NN (M-to-M or S-to-M) is more simple and light-weighted, denoted as $Enc_{\mathrm{sim}}$ and $Dec_{\mathrm{sim}}$ (combination denoted as $AE_{\mathrm{sim}}$).

\par Tab. \ref{Tab:complexity} compares the memory consumption, presented by the parameter amount $\sf{para}(\cdot)$, and the overall training and online testing time complexity, presented by the FLOPs amount $\sf{FLOPs}(\cdot)$. When comparing S-to-S with the multitasking modes (M-to-M and S-to-M ), the experiment results show that both the parameters and FLOPs amount of $AE_{\mathrm{com}}$ are much larger than $AE_{\mathrm{sim}}$, even larger than $T\cdot \mathsf{para}(AE_{\mathrm{sim}})$ or $T\cdot \mathsf{FLOPs}(AE_{\mathrm{sim}})$, demonstrating that even if in the training period, the time complexity of S-to-M’s joint architecture (joint training the shared encoder and multiple decoders) is lower than S-to-S.  
We then compare our proposed S-to-M mode with the M-to-M mode. As an upgraded version of M-to-M, S-to-M only saves one set of the encoder's parameters, reducing much memory consumption at the UE. 
In terms of complexity, S-to-M increases the complexity brought by the GateNet compared with M-to-M. However, since GateNet is a small network consisting of only three FC layers, this increased complexity can be ignored. Moreover, GateNet eliminates the need for the UE to provide additional information to the BS in real-time to switch network parameters.

\begin{table*}[t]
	\centering
	\setlength{\abovecaptionskip}{0cm}
    \setlength{\belowcaptionskip}{-0.2cm}
	\caption{\label{Tab:complexity} Comparison of memory consumption, offline training and online testing time complexity between S-to-S/M-to-M/S-to-M.}  
	\resizebox{1\textwidth}{!}{
	\begin{threeparttable}
	\begin{tabular}{c|cccc}
		\hline \hline
Mode & S-to-S & M-to-M & S-to-M + GateNet&  \\\hline
Enc Memory     & $\mathsf{para}(Enc_{\mathrm{com}})$ & $T \cdot \mathsf{para}(Enc_{\mathrm{sim}})$\tnote{1} & $\mathsf{para}(Enc_{\mathrm{sim}})$   \\
Dec Memory     & $\mathsf{para}(Dec_{\mathrm{com}})$ & $T \cdot \mathsf{para}(Dec_{\mathrm{sim}})$ & $T \cdot \mathsf{para}(Dec_{\mathrm{sim}})+\mathsf{para}(GateNet)$   \\
Overall training time complexity\tnote{2}     & $ \sum_{k=1}^{T}\left (N_k\cdot \mathsf{FLOPs}(AE_{\mathrm{com}})\right )$ & $\sum_{k=1}^{T}\left (N_k\cdot  \mathsf{FLOPs}(AE_{\mathrm{sim}})\right ) $ & $\sum_{k=1}^{T}\left ( N_k\cdot  (\mathsf{FLOPs}(AE_{\mathrm{sim}})+\mathsf{FLOPs}(GateNet))\right ) $       \\
Online testing time complexity   & $\mathsf{FLOPs}(AE_{\mathrm{com}})$ & $\mathsf{FLOPs}(AE_{\mathrm{sim}})$ & $\mathsf{FLOPs}(AE_{\mathrm{sim}})+\mathsf{FLOPs}(GateNet)$ 
\\\hline \hline
	\end{tabular}
	
	\begin{tablenotes}
       \footnotesize
       \item[1] Assuming that multiple networks have the same architecture is a convenient way to simplify the analysis.
       \item[2]  In the S-to-S mode, there is only one NN trained with $\sum_{k=1}^{T}N_k$ samples, which is same with S-to-M. However, in the M-to-M mode, there are $T$ small NNs, each trained with $N_k$ samples. Therefore, a total of $\sum_{k=1}^{T}N_k$ samples are needed for training.
     \end{tablenotes}
    \end{threeparttable}}
\vspace{-1cm} 
\end{table*} 
Due to the structure of multiple parallel decoders and the joint training process in S-to-M, the only drawback is that it may consume a significant amount of graphics memory during training when the sub-task number becomes too large. However, this issue is not a challenge for a BS equipped with high-performance equipment, and it can be resolved through parallel or distributed training with multiple GPUs.

\section{Simulation Results and Discussions}
\label{Simulation-Results}
First, we provide the CSI dataset simulation and NN training details.
Then, we evaluate and compare the proposed MTL-based CSI feedback deployment mode, S-to-M, with the benchmarks, S-to-S/M-to-M. Finally, we analyze the results and investigate the mechanism of S-to-M mode via the compressed code’s visualization.

\subsection{CSI Simulation Setting}
\subsubsection{QuaDRiGa simulation datasets}
\label{subsubsec:Qua-dataset}
The multi-scenarios CSI dataset is simulated using a quasi-deterministic radio channel model, with the QuaDRiGa software tool\cite{QuaDRiGa}\footnote{QuaDRiGa simulation software can be download at \href{https://quadriga-channel-model.de}{https://quadriga-channel-model.de}}. This software allows for the modeling of MIMO radio channels for various specific network configurations and is widely used in academic research. Basic information about the CSI simulation is provided in Tab. \ref{QuaDRiGaSetting}. To ensure that the CSI sampled within each subregion retains its WSS properties in LSPs, we restricted the range of each subregion to not exceed the spatial correlation distance. We simulate five typical environments in an urban cell, with the BS located at the center (0,0) and the UE randomly located in a subregion centered on marked coordinates. The scenario settings are provided in Tab \ref{5P-Setting}, and default settings according to \cite{QuaDRiGa} are used for the rest of the simulation. The visualization of the CSI maps and the sample correlation are shown in Figs \ref{fig:5P-graydraw} and \ref{fig:correlation_5P}, respectively, with the analysis presented in Section \ref{system model}.

\begin{table*}[t]
    \centering

    \begin{minipage}[t]{0.47\textwidth}
    \setlength{\abovecaptionskip}{0cm}
    \setlength{\belowcaptionskip}{-0.2cm}
    \centering
    \caption{\label{QuaDRiGaSetting}Basic parameter setting in the channel generation using QuaDRiGa software.}

	\renewcommand\arraystretch{0.65}
	\resizebox{0.98\textwidth}{!}{
	
	\begin{tabular}{l|l}
		\hline \hline
		\multirow{2}{*}{Antenna setting}&  32 ULA antennas at BS\\& single antenna at UE           \\ \hline
		\multirow{2}{*}{Operating system} & FDD-OFDM system \\& with 512 subcarriers\\ \hline
		Center frequency & 2.655GHz\\ 
		Bandwidth & 10MHz\\ \hline
		\multirow{2}{*}{Scenarios} &  3GPP-38.901-UMi-NLOS    \\
		&  3GPP-38.901-UMi-LOS \\\hline
	    Space correlation distant  &  20m\\
	    Scattering clusters number &  5/10/40\\	\hline
		Cell range        & $400m\times 400m$  \\
		Sub-area range         & $20m\times 20m$ \\\hline

	    Sampling number in each sub-area& 50,000\\
	    Sub-area's number$/$Tasks number &  5 \\\hline
	    \multirow{3}{*}{CSI Pretreatment} & 2D-DFT\\ & sparse clipping (reserve nonzero 32 rows)\\& normalized to real values in the range $[0,1]$ \\
		\hline \hline
	\end{tabular}}
    \end{minipage}
    \begin{minipage}[t]{0.47\textwidth}
    \centering
    \setlength{\abovecaptionskip}{0cm}
    \setlength{\belowcaptionskip}{-0.2cm}
    \caption{\label{5P-Setting}Differences in simulation dataset between sub-tasks.}
    \setlength{\tabcolsep}{4pt}
    \renewcommand\arraystretch{1.8}
    \resizebox{0.98\textwidth}{!}{
    \begin{tabular}{l|llll}
    	\hline \hline
    	Task&  Sub-area& Center position& Scenario& Clusters\\\hline
    	task1& Commercial area& (50,0)& 3GPP-38.901-UMi-NLOS& 40  \\ 
    	task2& Residential area& (-100,-50)& 3GPP-38.901-UMi-NLOS& 40  \\
    	task3& Park& (10, -70)& 3GPP-38.901-UMi-LOS& 5  \\
    	task4& Parking lot& (90, -160)& 3GPP-38.901-UMi-LOS& 5  \\
    	task5& Warehouse& (0, 170)& 3GPP-38.901-UMi-NLOS& 10  \\
    	\hline \hline
    \end{tabular}}
     \end{minipage}
\vspace{-0.4cm} 
\end{table*} 

\subsubsection{COST2100 Indoor/Outdoor datasets}
\begin{figure}[t]
	\centering 
	\setlength{\abovecaptionskip}{-0cm}
    \setlength{\belowcaptionskip}{-0.6cm}
	\label{fig:cost-visual}
	\subfigure [\label{fig:cost-graydraw}Gray scale image of CSI maps in the COST2100 Indoor/Outdoor datasets. The first row contains six samples randomly extracted from the Indoor dataset, while the samples in second row are from the Outdoor dataset.]{
		\includegraphics[width=0.95\linewidth]{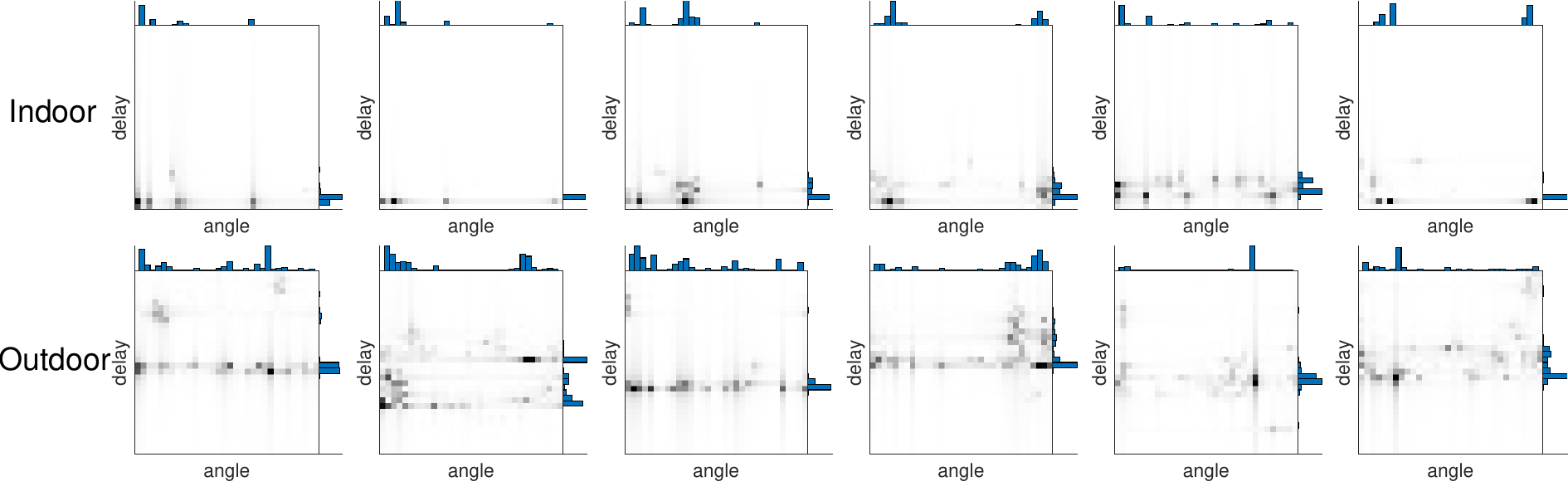}} 

	\subfigure [\label{fig:cost-correlation}Sample correlation matrix of PAS/PDP/CSI in the COST2100 dataset.]{
		\includegraphics[width=0.7\linewidth]{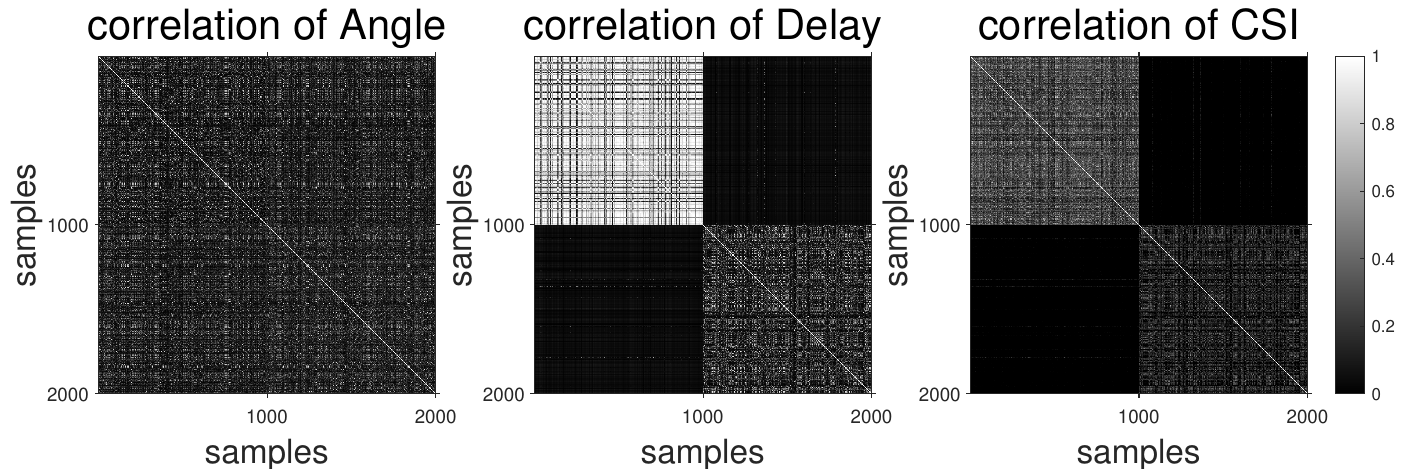}}
	\caption{Visualization of the COST2100 Indoor/Outdoor dataset.  } 
 	\vspace{-0.8cm} 
\end{figure}

We conduct experiments on the CSI dataset generated by the COST2100 channel model with Indoor/Outdoor scenarios to further verify the feasibility of our MTL-based mode. The dataset used in \cite{2018CsiNet} is employed, with the BS positioned at the center of a square area with lengths of 20m and 400m for indoor and outdoor scenarios, respectively, while the UEs are randomly placed in the square area per sample. We also analyze the CSI feature distribution of the COST2100 Indoor/Outdoor datasets. Fig. \ref{fig:cost-correlation} shows that the first 1000 samples were randomly taken from the Indoor dataset, and the last 1000 samples were taken from the Outdoor dataset. The difference in feature distribution mainly comes from the large difference in delay distribution between the two scenarios. In the Indoor scenario, the distance between UE and BS is within a range of 20 m, whereas the range enlarges to 200 m in the Outdoor scenario. There is no significant difference in the distribution of angular features because the sampling area does not involve any azimuth information (BS is centered in the sampling area). The sample correlation of the Indoor dataset is higher than that of the Outdoor dataset, which also explains why the same NN always performs better on the Indoor dataset than on the Outdoor dataset.

\subsection{NN architecture and training details}

\subsubsection{NN architecture of various complexity}
\begin{figure}
    \centering
    \setlength{\abovecaptionskip}{-0.2cm}
    \includegraphics[width=1\linewidth]{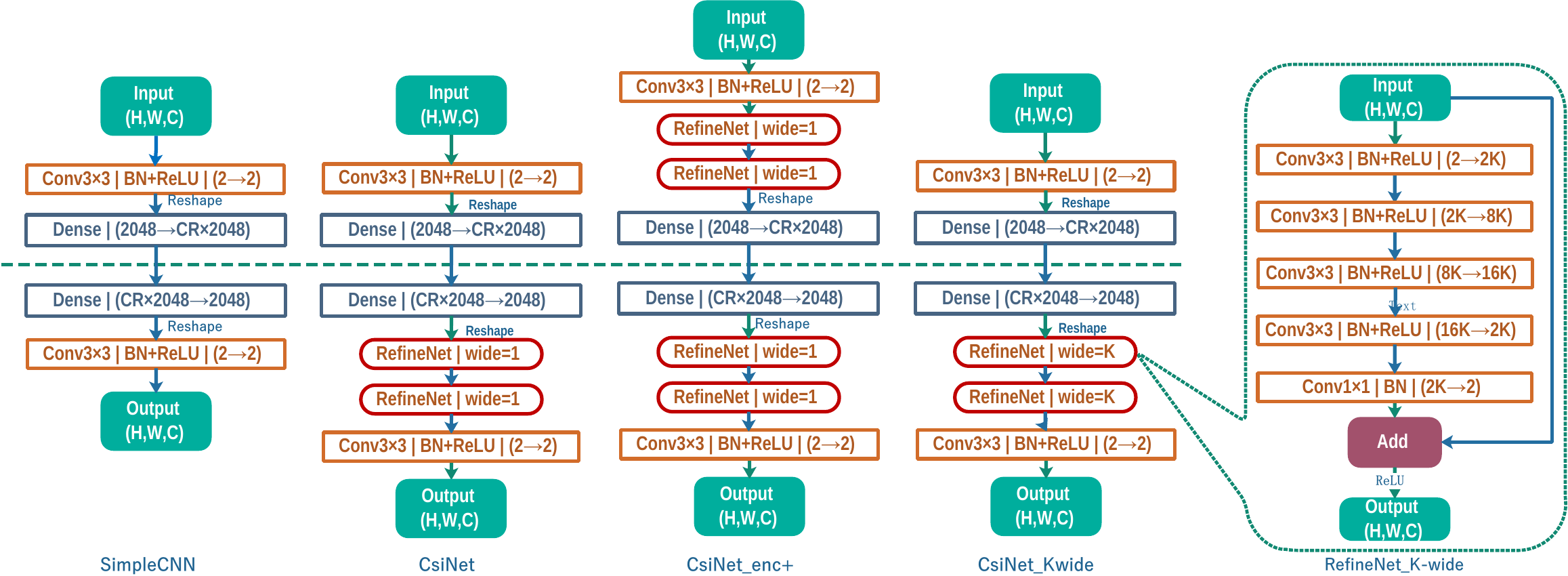}
    \caption{NN architectures with various complexities involved in the experiment based on CsiNet. The K-wide RefineNet is shown on the right side. In each layer, the details are marked as: Type and kernel size $|$ Batch Normalization and ReLU activation $|$ Input channels/Output channels.}
    \label{fig:NN-architecture}
\vspace{-0.8cm} 
\end{figure}

Note that any NN architecture can be used in the three modes that we proposed. In our experiments, we use NNs with varying complexities to validate our approach. For ease of comparison, we modify the NN architecture based on CsiNet \cite{2018CsiNet}, as shown in Fig. \ref{fig:NN-architecture}. The complexity of the NN is increased by widening the structure of the RefineNet (CsiNet\_K-wide) or adding additional RefineNet blocks in the encoder NN (CsiNet\_enc+). We measure the time and space complexity of the involved NNs using FLOPs and parameters, as shown in Tab. \ref{Tab:NN-complexity}. Tab. \ref{Tab:NN-complexity-specific} provides the specific complexity values of the three modes by combining Tab \ref{Tab:complexity} and \ref{Tab:NN-complexity}.

\begin{table}[t]
	\centering
	\setlength{\abovecaptionskip}{-0cm}
    \setlength{\tabcolsep}{3.5pt}
    \renewcommand\arraystretch{0.8}
	\caption{\label{Tab:NN-complexity} Time and space complexity of the involved NNs.}  
	\resizebox{0.85\textwidth}{!}{
	\begin{tabular}{c|ccccc|ccccc}
		\hline \hline
		\multirow{2}{*}{CR}&\multicolumn{5}{c|}{FLOPs(AE)[M]}&\multicolumn{5}{c}{Para(AE)[M]}\\
		  &SimpleCNN &CsiNet & CsiNet\_enc+ &  CsiNet\_8wide & CsiNet\_16wide &SimpleCNN &CsiNet & CsiNet\_enc+ &  CsiNet\_8wide & CsiNet\_16wide\\\hline
		
		$1/4$ & 2.17 &6.23 &10.30 &210.44 &833.95 & 2.10  &2.10  &2.11  &2.30  &2.91\\
		$1/8$ & 1.12 &5.19 &9.25 &209.40 &832.91 & 1.05 &1.05   &1.06 &1.25  &1.86\\
		$1/16$ & 0.60 &4.66 &8.72  &208.87  &832.38 & 0.53 &0.53 &0.53  &0.73   &1.34 \\
		$1/32$ & 0.34 &4.40 &8.46  &208.61  &832.12 & 0.27 &0.27 &0.27  &0.47   &1.08 \\
		$1/64$ & 0.20 &4.27 &8.33 &208.48 &831.99 & 0.14 &0.14 &0.14  &0.34  &0.95 \\\hline
		\hline \hline
	\end{tabular}}
\vspace{-0.8cm} 
\end{table}
\begin{table}[t]
	\centering
	\setlength{\abovecaptionskip}{-0cm}
    \setlength{\tabcolsep}{10.0pt}
    \renewcommand\arraystretch{0.8}
	\caption{\label{Tab:NN-complexity-specific} Time and space complexity of S2S/M2M/S2M with the involved NNs.}  
	\resizebox{0.95\textwidth}{!}{
     \begin{threeparttable}
	\begin{tabular}{c|ccc|ccc|ccc|ccc} \hline \hline
\multirow{2}{*}{CR=1/4} & \multicolumn{3}{c|}{\begin{tabular}[c]{@{}c@{}}Enc Memory\\ (Para{[}M{]})\end{tabular}} & \multicolumn{3}{c|}{\begin{tabular}[c]{@{}c@{}}Dec Memory\\ (Para{[}M{]})\end{tabular}} & \multicolumn{3}{c|}{\begin{tabular}[c]{@{}c@{}}Overall training time \\ complexity (FLOPs{[}T{]})\tnote{2}\end{tabular}} & \multicolumn{3}{c}{\begin{tabular}[c]{@{}c@{}}Online testing time \\ complexity (FLOPs{[}M{]})\end{tabular}} \\
& S2S  & M2M  & $\rm{S2M}^{*}$ \tnote{1}                    
& S2S  & M2M   & $\rm{S2M}^{*}$                     
& S2S & M2M   & $\rm{S2M}^{*}$  
& S2S & M2M   & $\rm{S2M}^{*}$   \\\hline
SimpleCNN  & 1.05   & 5.25   & 1.05  & 1.05                      & 5.25                     & 7.35                            & 0.543                             & 0.543                             & 1.065                                  & 2.17                             & 2.17                             & 4.26                                   \\
CsiNet                  & 1.05                      & 5.25                     & 1.05                            & 1.05                      & 5.25                     & 7.35                            & 1.558                             & 1.558                             & 2.080                                  & 6.23                             & 6.23                             & 8.32                                   \\
CsiNet\_enc+            & 1.05                      & 5.25                     & 1.05                            & 1.05                      & 5.25                     & 7.35                            & 2.58                              & 2.58                              & 3.098                                  & 10.3                             & 10.3                             & 12.39                                  \\
CsiNet\_8wide           & 1.05                      & 5.25                     & 1.05                            & 1.25                      & 6.25                     & 8.34                            & 52.60                             & 52.60                             & 53.13                                  & 210.4                            & 210.4                            & 212.5                                  \\
CsiNet\_16wide          & 1.05                      & 5.25                     & 1.05                            & 1.86                      & 9.30                     & 11.39                           & 208.5                             & 208.5                             & 209.0                                  & 833.9                            & 833.9                            & 836.0       \\\hline  \hline                        
\end{tabular}
    \begin{tablenotes}
       \footnotesize
       \item[1] $\rm{S2M}^{*}$ denotes S-to-S mode with GateNet.
       \item[2]  In calculation, the training samples N=50,000 for each subtask. The overall training time is measured by training samples $\times$ FLOPs of one sample, with the unit of trillion.
     \end{tablenotes}
    \end{threeparttable}
 }
 \vspace{-1cm}
\end{table}

\subsubsection{Training details}
During training with the QuaDRiGa simulation datasets, each subtask is composed of 50,000 samples. These samples are split into training, validation, and test sets, which account for 85\%, 10\%, and 5\% of the total number of samples, respectively. For training with the COST2100 dataset, both the Indoor and Outdoor scenarios have 100,000 samples for training, 30,000 samples for validation, and 20,000 samples for testing. 
For QuaDRiGa dataset, the total number of training samples for S-to-S/M-to-M/S-to-M is $250,000\times 85\%$, according to Note 2 of Tab. \ref{Tab:complexity}. For COST2100 Indoor/Outdoor dataset, the total number of training samples for S-to-S/M-to-M/S-to-M is $200,000$.
We use the Adam optimizer with a learning rate of 1e-3 and a training epoch of 1,200. To prevent overfitting, we adopt the early stopping method during training.

The supervised learning for the GateNet is carried out after the AE's training. One-hot labels corresponding to the task category are generated and paired with the encoder's output. All the paired samples, i.e., 250,000 QuaDRiGa samples and 200,000 COST2100 samples, are mixed up. The training, validation, and test data comprise 85\%, 10\%, and 5\%, respectively. The training is performed using Adam with a learning rate of 1e-3, and 300 epochs are trained for all experiments.

\subsection{Performance of MTL-based CSI feedback mode}
In this subsection, we evaluate and compare the proposed MTL-based CSI feedback deployment mode, S-to-M, with the benchmarks, S-to-S/M-to-M.

\subsubsection{Performance of the GateNet}
Tab. \ref{Tab:GateNet_performance} presents the performance of the GateNet, including its inference accuracy and the NMSE gap between the overall S-to-M (with GateNet) and the upper bound. The upper bound is obtained by substituting the GateNet output with the ground truth label. The NMSE gap measures the loss of NMSE performance resulting from the mismatch with the task-specific decoder caused by the wrong task number output by GateNet. In the first step of training, CsiNet is used to obtain the trained shared encoder. The GateNet shows rapid convergence to high accuracy, achieving approximately 99.9\% accuracy at 50 epochs. Even at an extremely low compression ratio (CR=1/64), the accuracy can still converge to 99.98\%, which has negligible effects on the S-to-M mode (the NMSE gap is less than 1e-2 dB). These results indicate that the compressed code contains sufficient valuable information for the classifier to identify its task category.

\begin{table*}[t]
    \centering
    \begin{minipage}[t]{0.45\textwidth}
    \centering
    \setlength{\abovecaptionskip}{0cm} 
	\setlength{\belowcaptionskip}{-0.2cm}
    \caption{\label{Tab:GateNet_performance} GateNet performance tested with CsiNet on two datasets.}
    
    \setlength{\tabcolsep}{3.5pt}
	\renewcommand\arraystretch{1.203}
	\resizebox{0.98\textwidth}{!}{
	\begin{tabular}{c|c|ccccc|cc}
		\hline \hline
		&\multirow{2}{*}{CR} &\multicolumn{5}{c|}{QuaDRiGa}&\multicolumn{2}{c}{COST2100}\\
		& &T1   &T2 &T3 &T4 &T5 &Indoor &Outdoor\\
		\hline
		\multirow{5}{*}{\begin{sideways}{Accuracy [\%]}\end{sideways}}
		&$1/4$ &99.996 & 99.994 &100.00 &100.00 &100.00    &99.999 &100.00 \\
		&$1/8$ &100.00 & 100.00 &100.00 &100.00 &100.00    &99.999 &100.00 \\
		&$1/16$ &99.996 & 99.952 &99.944 &99.980 &99.992    &100.00 &100.00  \\
		&$1/32$ &100.00 & 99.988 &99.968 &99.992 &99.996    &99.997 &100.00  \\
		&$1/64$ &100.00 & 99.984 &99.960 &99.994 &99.910    &100.00 &100.00  \\\hline
        \multirow{5}{*}{\begin{sideways}{NMSE Gap [dB]}\end{sideways}}
		&$1/4$ &1.80e-03	&4.93e-03 &	0.00e+00	&0.00e+00	&0.00e+00	&1.04e-03	&0.00e+00
 \\
		&$1/8$ &0.00e+00	&0.00e+00	&0.00e+00	&0.00e+00	&0.00e+00	&9.67e-04	&0.00e+00
  \\
		&$1/16$ &6.72e-03	&1.47e-02	&6.08e-02	&6.32e-03	&4.15e-03	&0.00e+00	&0.00e+00
   \\
		&$1/32$ &0.00e+00	&2.54e-03	&2.87e-02	&1.22e-03	&1.41e-03	&7.29e-04	&0.00e+00
   \\
		&$1/64$ &0.00e+00	&2.88e-03	&3.33e-02	&1.54e-03	&3.90e-02	&0.00e+00	&0.00e+00
   \\

		\hline \hline
	\end{tabular}}
    \end{minipage}
    \begin{minipage}[t]{0.45\textwidth}
    \centering
    \setlength{\abovecaptionskip}{0cm} 
	\setlength{\belowcaptionskip}{-0.2cm}
    \caption{\label{Tab:5P-NN-performance} NMSE performance of the overall S-to-M mode tested with CsiNet.}
    \renewcommand\arraystretch{0.83}
    \resizebox{0.98\textwidth}{!}{
    \begin{tabular}{c|c|ccccc|cc}
		\hline \hline
		\multirow{2}{*}{CR} &\multirow{2}{*}{Mode} &\multicolumn{5}{c|}{QuaDRiGa} & \multicolumn{2}{c}{Cost2100} \\ 
		                    &                      &\multicolumn{1}{c}{Task1}&\multicolumn{1}{c}{Task2}&\multicolumn{1}{c}{Task3}&\multicolumn{1}{c}{Task4}&Task5&\multicolumn{1}{c}{Indoor}&\multicolumn{1}{c}{Outdoor} \\ \hline
		\multirow{3}{*}{1/4}& S-to-S &-7.96 &-12.24 &-14.71 &-12.61  &-12.07 &-11.74 	&-8.54 \\
		                    & M-to-M &-8.06         &\textbf{-15.85} &\textbf{-19.48} &\textbf{-16.76} &-14.61 &-15.00 	&-8.59 \\
		                    & S-to-M &\textbf{-9.26} &	-15.39 &	-18.80          &	-16.04 &	\textbf{-14.94} &\textbf{-15.25} &	\textbf{-9.00} \\ \hline
		\multirow{3}{*}{1/8}& S-to-S &-5.79 &	-9.44 &	-13.31 &	-10.99 &	-10.88  &-10.84 &	-6.44 \\
		                    & M-to-M &-5.61 &	-10.11 &	\textbf{-18.20} &	\textbf{-13.71} 	&-12.71  &-11.22 &	-6.34 \\
		                    & S-to-M &\textbf{-6.63} &	\textbf{-11.65} &	-17.90 &	-13.49 &\textbf{-13.29}  &\textbf{-13.08} &	\textbf{-6.78} \\ \hline    
		\multirow{3}{*}{1/16}& S-to-S &-2.78 &	-4.95 &	-8.17 &	-6.22 &	-6.84   &-7.23 	&-3.82  \\
		                    & M-to-M &\textbf{-4.32} &	\textbf{-7.55} &	\textbf{-14.20} 	&\textbf{-11.45} &	-9.14   &-8.24 &	-4.02 \\
		                    & S-to-M &-3.85 &	-6.64 &	-11.83 &	-8.87 &	\textbf{-9.24}   &\textbf{-8.39} &	\textbf{-4.17}  \\ \hline  
		\multirow{3}{*}{1/32}& S-to-S &-2.13 &	-3.11 &	-5.83 &	-4.51 &	-4.83    &-6.98	& -2.96  \\
		                    & M-to-M &\textbf{-3.61} &	\textbf{-4.64} &	\textbf{-14.14} &	\textbf{-8.87} &	\textbf{-9.14}    &-7.15	& \textbf{-3.09} \\
		                    & S-to-M &-3.17 &	-4.39 &	-10.68 &	-7.24 &	-7.72    &\textbf{-7.95}	& \textbf{-3.09} \\ \hline  		                    
		\multirow{3}{*}{1/64}& S-to-S &-2.64 &	-2.98 &	-4.16 &	-5.92 &	-6.14     &-4.50 &	-1.58  \\
		                    & M-to-M &\textbf{-2.95} &	-3.21 &	\textbf{-10.03} &	-6.74 &	\textbf{-7.02}     &\textbf{-4.70} &	-1.65 \\
		                    & S-to-M &-2.91 &	\textbf{-3.29} &	-9.56 &	\textbf{-7.09} &	-7.00     &-4.00 &	\textbf{-1.69} \\  
		\hline \hline
	\end{tabular}}
     \end{minipage}
\vspace{-0.4cm} 
\end{table*}
\subsubsection{Feasibility of S-to-M mode}
\begin{figure}
    \centering
    \setlength{\abovecaptionskip}{-5mm}
    \setlength{\belowcaptionskip}{0.cm}
    \includegraphics[width=\textwidth]{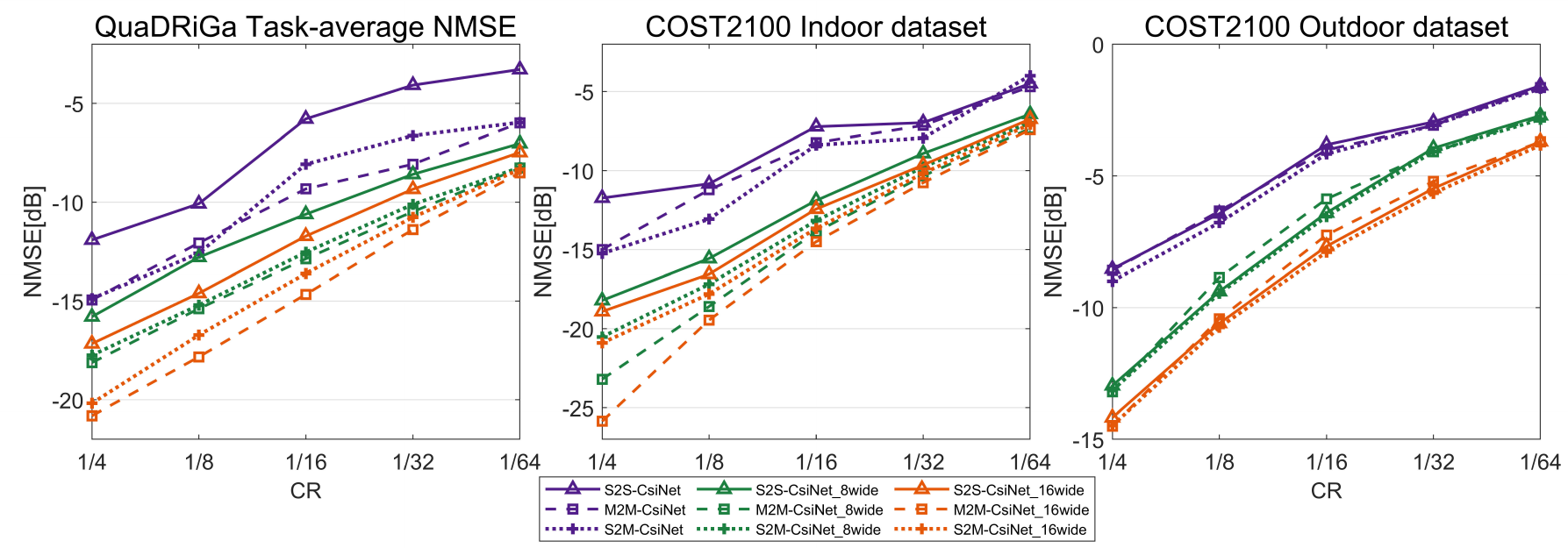}
    \caption{NMSE Performance of the S-to-S/M-to-M/S-to-M mode of various NN and CR trained on the QuaDRiGa simulation datasets (left subgraph) or the COST2100 Indoor/Outdoor datasets (middle and right subgraphs).}
    \label{fig:performance-lines}
    \vspace{-0.8cm}
\end{figure}
We conduct experiments on simulation datasets generated by QuaDRiGa and the COST2100 Indoor/Outdoor datasets. The performance of the CsiNet NN with various compression ratios ($\mathrm{CR}=\frac{dim(\mathbf{s})}{2N_{\mathrm{c}}N_{\mathrm{t}}}$, where $dim(\mathbf{s})$ denotes the length of the feedback code) is evaluated on both datasets and the results are shown in Tab. \ref{Tab:5P-NN-performance}, where the best performance for each subtask is highlighted in bold. 
Fig. \ref{fig:performance-lines} shows the NMSE performance of various NNs and compression ratios trained using the S-to-S, M-to-M, and S-to-M modes on the QuaDRiGa simulation datasets (left subgraph) and the COST2100 Indoor/Outdoor datasets (middle and right subgraphs). The different complexities of the three NNs are indicated by different colors in the legend. The S-to-S, M-to-M, and S-to-M modes are represented by solid, long-dotted, and short-dotted lines, respectively. As shown in Fig. \ref{fig:performance-lines}, there is a trend of improving NMSE performance as the NN complexity increases. The dotted line is below the real line of the same color, revealing that S-to-M outperforms than S-to-S. It is important to note that the average NMSE performance of the five subtasks is shown in Fig. \ref{fig:performance-lines} for convenience, while the detailed performance differences between different subtasks are presented in Tab. \ref{Tab:5P-NN-performance}.

\begin{figure}
    \centering
    \setlength{\abovecaptionskip}{0mm}
    \setlength{\belowcaptionskip}{0.cm}
    \subfigcapskip=-5mm
    \subfigbottomskip=0mm
	\subfigure[Training on the QuaDRiGa simulation dataset\label{5P-NN-performance}]
     {\includegraphics[width=0.9\linewidth]{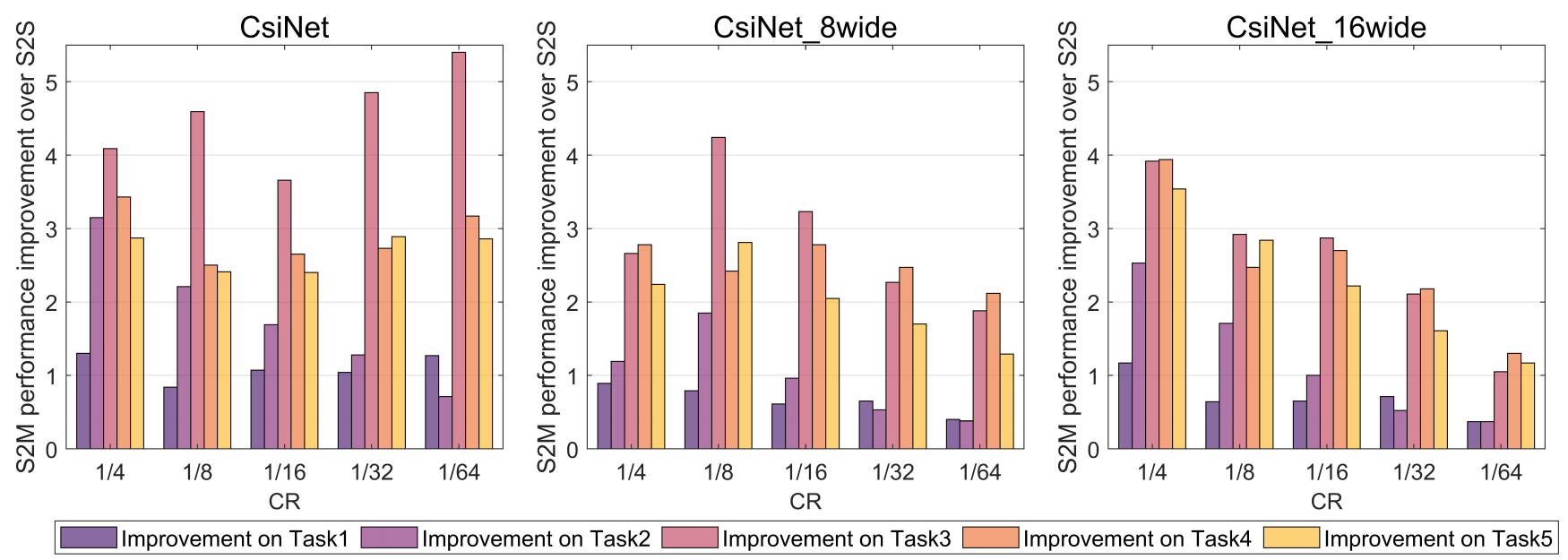}}
    \subfigure[Training on the COST2100 Indoor/Outdoor datasets\label{2P-NN-performance}]{\includegraphics[width=0.9\linewidth]{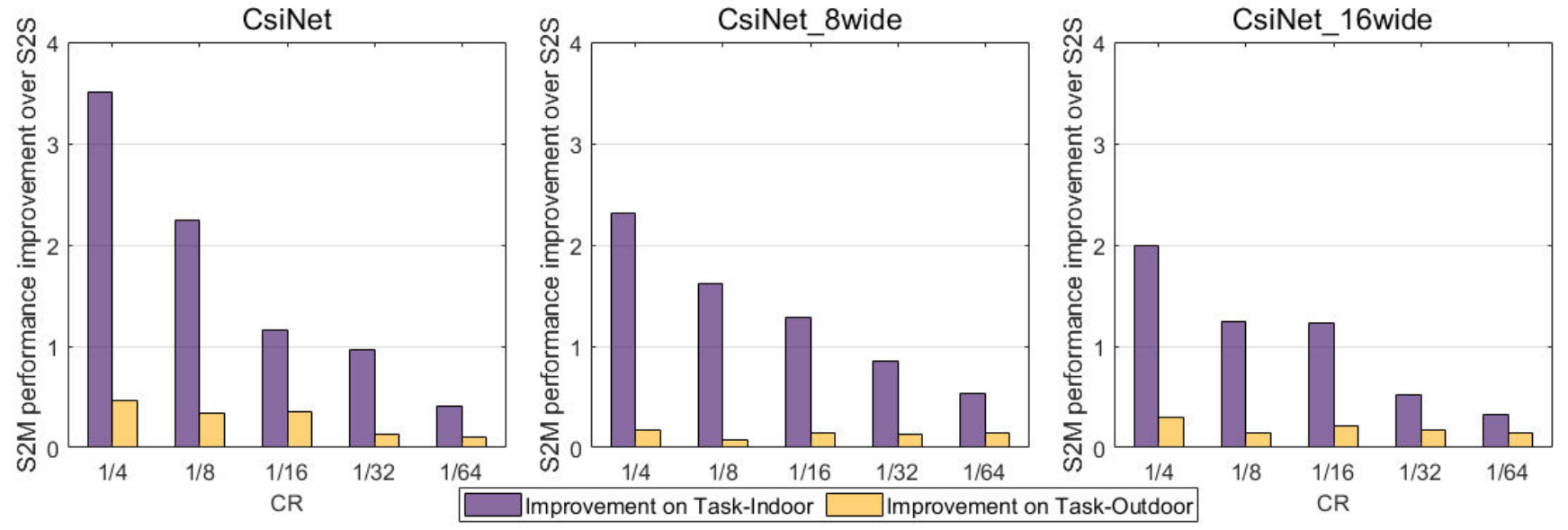}}
    \caption{NMSE improvement/gap of S-to-M over S-to-S mode for various NN and CR trained on the QuaDRiGa simulation datasets (a) or the COST2100 Indoor/Outdoor datasets (b).}     
    \label{fig:NN-gap}
    \vspace{-0.8cm} 
\end{figure}

Fig. \ref{fig:NN-gap} shows the specific NMSE gap between S-to-M and S-to-S, where S-to-M outperforms S-to-S. The first and second rows present results for the QuaDRiGa and COST2100 datasets, respectively, and three NNs of different complexities are tested in each row from left to right. In each subgraph, the encoder and decoder NN architectures for S-to-M and S-to-S are the same ($AE_{com}$ equals $AE_{sim}$ in Tab. \ref{Tab:complexity}). Consequently, the sum parameters of all task-specific decoders of S-to-M are $T$ times that of S-to-S, but the BS has sufficient memory and capacity to store multiple sets of task-specific decoder parameters. Ignoring the slight complexity of GateNet (in Tab. \ref{Tab:complexity}), the overall training time complexity and online testing time complexity of S-to-M and S-to-S are equivalent. The bar in the figure displays the NMSE gap of S-to-M performance over S-to-S, i.e., $\mathrm{NMSE}(\mathrm{S-to-S})-\mathrm{NMSE}(\mathrm{S-to-M})$ $\mathrm{[dB]}$. All five subtasks are color-coded, and a positive bar value indicates S-to-M outperforms S-to-S, with the overperformance amplitude varying with the task.

In summary, several conclusions can be drawn from the results:
\\(a) When comparing the same NN structure, the S-to-M and M-to-M modes (represented by the short-dotted and long-dotted lines in Fig. \ref{fig:performance-lines}) perform better than the S-to-S mode. This result is due to the decomposition of a complex task into several easier subtasks, which has a positive effect on overall performance. This result is also reflected in the positive bar values in Fig. \ref{fig:NN-gap}, where the NMSE of S-to-M is lower than that of S-to-S.
\\
(b) Among the multitasking modes, S-to-M can achieve similar or even better performance than M-to-M with fewer sets of encoder parameters. This result is due to the hard sharing mechanism in MTL, which allows the shared encoder to access all subtask samples and thereby enhances the data. As a result, the risk of overfitting is reduced, and a smaller encoder can achieve the same or better generalization capability than a more complex encoder.

Some secondary observations can be made regarding the differences in performance for different subtasks, CR, or NN complexity. Firstly, S-to-M outperforms S-to-S more significantly when a simpler NN is used, as reflected in Fig. \ref{5P-NN-performance}. Additionally, the NMSE gap between S-to-M/M-to-M and S-to-S is greater in easier subtasks, such as QuaDRiGa subtask3 or COST2100 task-Indoor, compared to harder subtasks, like QuaDRiGa subtask1 or COST2100 task-Outdoor, as shown in Tab. \ref{Tab:5P-NN-performance} and Fig. \ref{2P-NN-performance}, respectively. Multitasking modes may not be as beneficial for networks with high complexity since the subtask's data samples may not be sufficient to train a large-capacity NN, which can lead to overfitting. This issue can be addressed by increasing the amount of training data or improving data quality.

The experimental results obtained on the COST2100 dataset are consistent with those on the QuaDRiGa dataset, with the exception of the Outdoor subtask. In this subtask, the multitasking modes do not show any significant advantage over S-to-S, as the NMSE performance lines of the three modes are almost overlapping, and the performance improvement (blue bar) is minimal. This phenomenon can be explained from the perspective of transfer learning. The small sampling range and high sample similarity of the Indoor task make it an easy-handled subtask, where 100,000 training samples are sufficient for the NN to grasp the feature representation. However, mixing training with other samples with large differences only leads to a performance loss of the network in the current task. Thus, Outdoor samples bring a negative transfer effect to the Indoor task. In this situation, M-to-M is a better selection for the Indoor task than S-to-S, and then the MTL-based improved version, S-to-M, can be adopted for the convenience of UE. For the Outdoor task, the sampling range is much larger than that of the Indoor task, and 100,000 samples are not enough for training, which explains why the NN always obtains worse performance than on the Indoor task. Due to the large differences between the Indoor/Outdoor samples, adding additional Indoor samples hardly provides any valuable assistance to the NN to fit the feature distribution of the Outdoor samples. Thus, the performances of the three modes on the Outdoor task are almost identical. To avoid formidable tasks like the Outdoor task, we can appropriately narrow the sampling range to improve the sample similarity by further segmenting the 200 m $\times$ 200 m Outdoor sampling place and decomposing this complex task into several easy-handling sub-tasks, where our S-to-M mode can bring noticeable performance enhancement.
 
\subsubsection{Advantages of S-to-M mode}
\begin{figure}
    \centering
    \setlength{\abovecaptionskip}{-5mm}
    \setlength{\belowcaptionskip}{0.cm}
    \includegraphics[width=0.8\linewidth]{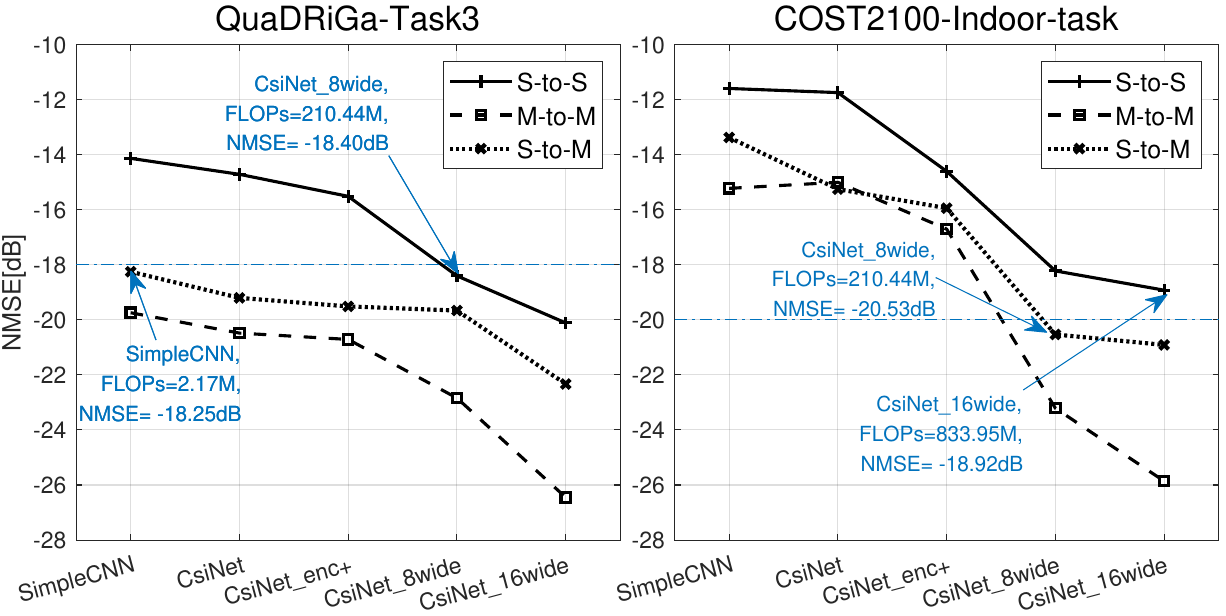}
    \caption{Model Complexity versus Performance in S-to-S/S-to-M/M-to-M mode on the QuaDRiGa-Task3 (left) or the COST2100 Indoor task (right) when CR=1/4.  
}
    \label{fig:comp-vs-perf}
\vspace{-1cm} 
\end{figure}
To ensure fairness in the comparison between S-to-M and S-to-S, additional experimental results are presented in Fig. \ref{fig:comp-vs-perf} to demonstrate the trade-off between NN complexity and performance. The left subgraph shows the result of QuaDRiGa Task3, while the right subgraph shows the result of the COST2100 Indoor task. The models marked in each subgraph achieve similar performance. In the QuaDRiGa Task3's result, S-to-S needs to increase the FLOPs of NN from SimpleCNN's 2.17 M to CsiNet\_8wide's 210.44 M (nearly 100 times) to achieve similar performance with S-to-M. In the COST2100 Indoor task result, S-to-M achieves better NMSE performance than -20 dB with a 210.44 M FLOPs NN, while S-to-S's performance cannot reach -20 dB even with a more complex 833.95 M FLOPs NN (more than two times). These results suggest that it is not cost-effective to increase the complexity of the S-to-S network to match the NMSE performance of S-to-M, as the performance of NMSE may not exceed that of S-to-M, even if the amplification multiple is far more than the number of tasks $T$.

The results suggest that using the region-specific S-to-M mode proposed in the paper yields higher benefits than the complex S-to-S network for learning the global cell's CSI. On the one hand, S-to-M can greatly reduce the network complexity while achieving the same level of reconstruction accuracy as S-to-S. On the other hand, S-to-M also achieves higher reconstruction accuracy under the same network complexity as S-to-S. When comparing the second case of using multiple region-specific and relatively simple NNs in the M-to-M mode, the advantages of S-to-M become more apparent. Tab. \ref{Tab:MM-NN-performance} shows that S-to-M can greatly reduce encoder memory consumption to $1/T$ of M-to-M and still achieve performance as good as M-to-M, owing to the hard sharing mechanism of MTL.


\begin{table}[t]
	\centering
	\setlength{\abovecaptionskip}{0cm} 
	\setlength{\belowcaptionskip}{-0.2cm}
	\caption{\label{Tab:MM-NN-performance} NMSE performance and encoder's complexity of M-to-M/ S-to-M tested of CsiNet.}  
	\renewcommand\arraystretch{0.8}
	\resizebox{0.8\textwidth}{!}{
	\begin{tabular}{c|c|ccccccc|ccccccc}
		\hline \hline
		\multirow{3}{*}{CR} &\multirow{3}{*}{Mode} &\multicolumn{7}{c|}{CsiNet\_8wide} & \multicolumn{7}{c}{CsiNet\_enc+} \\
		                    &                      &\multicolumn{5}{c|}{NMSE[dB]} & \multicolumn{2}{c|}{Enc\_complexity}    &\multicolumn{5}{c|}{NMSE[dB]} & \multicolumn{2}{c}{Enc\_complexity}\\
		                    &                      &\multicolumn{1}{c}{Task1}&\multicolumn{1}{c}{Task2}&\multicolumn{1}{c}{Task3}&\multicolumn{1}{c}{Task4}&\multicolumn{1}{c|}{Task5}&\multicolumn{1}{c}{para[M]}&\multicolumn{1}{c|}{FLOPs[M]} &\multicolumn{1}{c}{Task1}&\multicolumn{1}{c}{Task2}&\multicolumn{1}{c}{Task3}&\multicolumn{1}{c}{Task4}&\multicolumn{1}{c|}{Task5}&\multicolumn{1}{c}{para[M]}&\multicolumn{1}{c}{FLOPs[M]} \\ \hline
		\multirow{2}{*}{1/4}& M-to-M &-10.81 	&-16.23 	&-22.84 	&-21.35 	&-17.70         &\multicolumn{1}{|c}{5.25}&1.09
		                             &-8.46 	&-15.95 	&-19.87	&-17.12 	&-14.88         &\multicolumn{1}{|c}{5.25}&5.15\\
		                    & S-to-M &-11.80 	&-16.47 	&-19.66 	&-20.23 	&-19.70    &\multicolumn{1}{|c}{1.05}&1.09
		                             &-9.56 	&-15.69 	&-18.53 	&-16.54 	&-15.13   &\multicolumn{1}{|c}{1.05}&5.15\\ \hline
		                             
		\multirow{2}{*}{1/8}& M-to-M &-8.30 	&-12.93 	&-21.61 	&-18.82 	&-15.30    &\multicolumn{1}{|c}{2.625}&0.516
		                            &-5.77 	&-10.45 	&-18.89 	&-14.11 	&-12.91     &\multicolumn{1}{|c}{2.640}&4.620\\
		                    & S-to-M &-8.11 	&-13.37 	&-20.28 	&-17.86 	&-16.45   &\multicolumn{1}{|c}{0.525}&0.516
		                            &-6.83 	&-11.95 	&-18.20 	&-13.76 	&-13.55    &\multicolumn{1}{|c}{0.528}&4.620\\ \hline    
		                            
		\multirow{2}{*}{1/16}& M-to-M &-5.56 	&-8.92 	&-19.81 	&-16.93 	&-13.14     &\multicolumn{1}{|c}{1.310}&0.299
		                            &-4.52 	&-7.78 	&-14.56 	&-11.79 	&-9.53    &\multicolumn{1}{|c}{1.330}&4.360\\
		                    & S-to-M &-5.97 	&-9.16 	&-17.62 	&-16.12 	&-13.84      &\multicolumn{1}{|c}{0.262}&0.299
		                              &-3.95 	&-6.88 	&-12.54 	&-9.87 	&-9.55 &\multicolumn{1}{|c}{0.266}&4.360\\ 
		                              \hline  
		                              
		\multirow{2}{*}{1/32}& M-to-M &-4.35 	&-5.70 	&-15.73 	&-15.97 	&-10.67      &\multicolumn{1}{|c}{0.656}&0.168
		                            &-3.65 	&-4.69 	&-13.04 	&-8.89 	&-9.12       &\multicolumn{1}{|c}{0.672}&4.23\\
		                    & S-to-M &-4.65 	&-5.90 	&-14.06 	&-14.11 	&-11.56     &\multicolumn{1}{|c}{0.131}&0.168
		                            &-3.27 	&-4.54 	&-11.21 	&-7.87 	&-8.12      &\multicolumn{1}{|c}{0.134}&4.23\\ 
		                            \hline  
		                            
		\multirow{2}{*}{1/64}& M-to-M &-3.86 	&-4.14 	&-11.13 	&-13.34 	&-8.91       &\multicolumn{1}{|c}{0.328}&0.102
		                            &-2.95 	&-3.20 	&-10.12 	&-6.64 	&-7.05       &\multicolumn{1}{|c}{0.345}&4.17\\
		                    & S-to-M &-3.76 	&-4.32 	&-10.81 	&-12.94 	&-9.33       &\multicolumn{1}{|c}{0.066}&0.102
		                            &-2.91 	&-3.29 	&-9.78 	&-7.11 	&-7.02       &\multicolumn{1}{|c}{0.069}&4.17\\  
		\hline \hline
	\end{tabular}}
	\vspace{-0.8cm} 
\end{table}

\subsubsection{Interpretability of the shared encoder's mechanism in S-to-M}\label{Interpretability}
In the architecture of an AE, the encoder's role is to compress the input CSI matrix into an efficient data representation, and its effectiveness is determined by the accuracy of the decoder's reconstruction of the compressed coding. In the case of S-to-M and S-to-S, both have access to all the sub-task data via the same encoder NN structure, but the encoder of S-to-M outperforms that of S-to-S. To better understand the mechanism of S-to-M mode, we investigate the unsupervised learning of the AE through the visualization of the compressed code. Specifically, a good encoder can map the high-dimensional CSI data to a low-dimensional embedding space, such that the distribution of sample points in the embedding space is sufficiently dispersed, and the difference between sample points is obvious, allowing the decoder to distinguish between samples during reconstruction. If two disparate CSI samples are encoded into overlapping embedding points, the decoder may reconstruct them into very similar or even identical matrices. The decoder, in turn, acts as a generator that generates the original high-dimensional CSI from the information contained in the embedding vectors.

\begin{figure}
    \centering
    \setlength{\abovecaptionskip}{0mm}
    \setlength{\belowcaptionskip}{0.cm}
	\subfigure[S-to-S\label{5P_simple_mixcode}]{\includegraphics[width=0.3\linewidth]{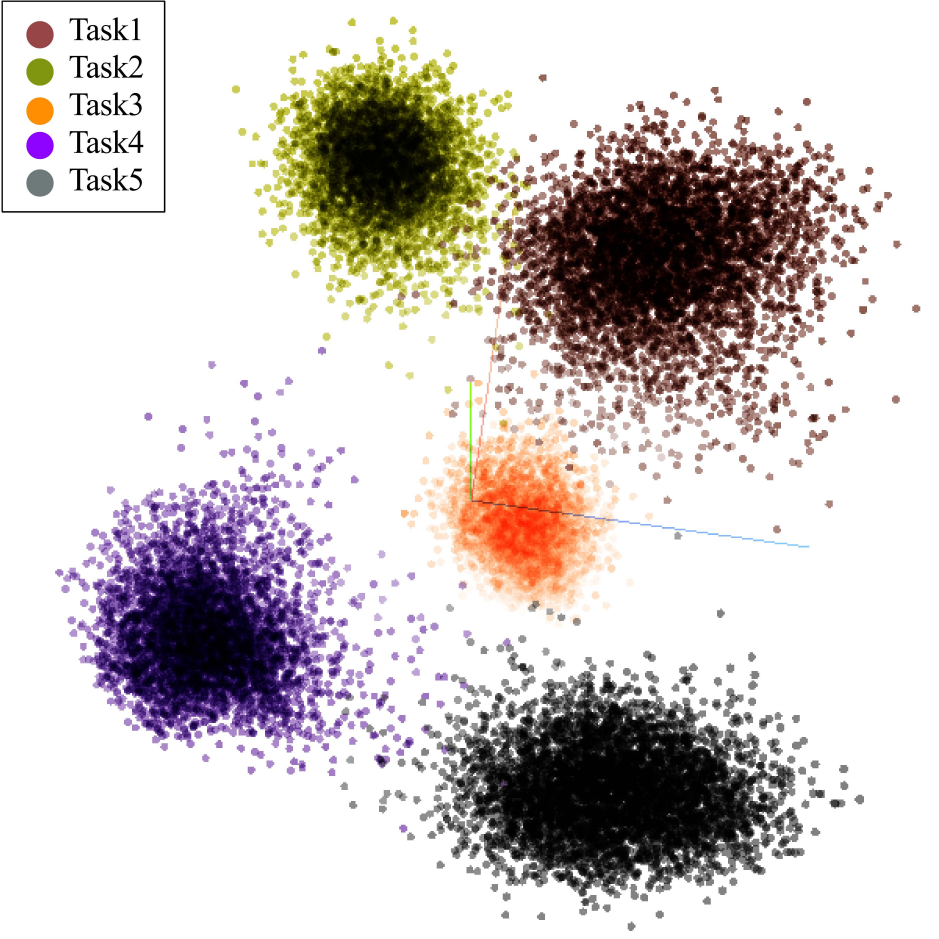}}
    \subfigure[S-to-M\label{5P_simple_MTLcode}]{\includegraphics[width=0.3\linewidth]{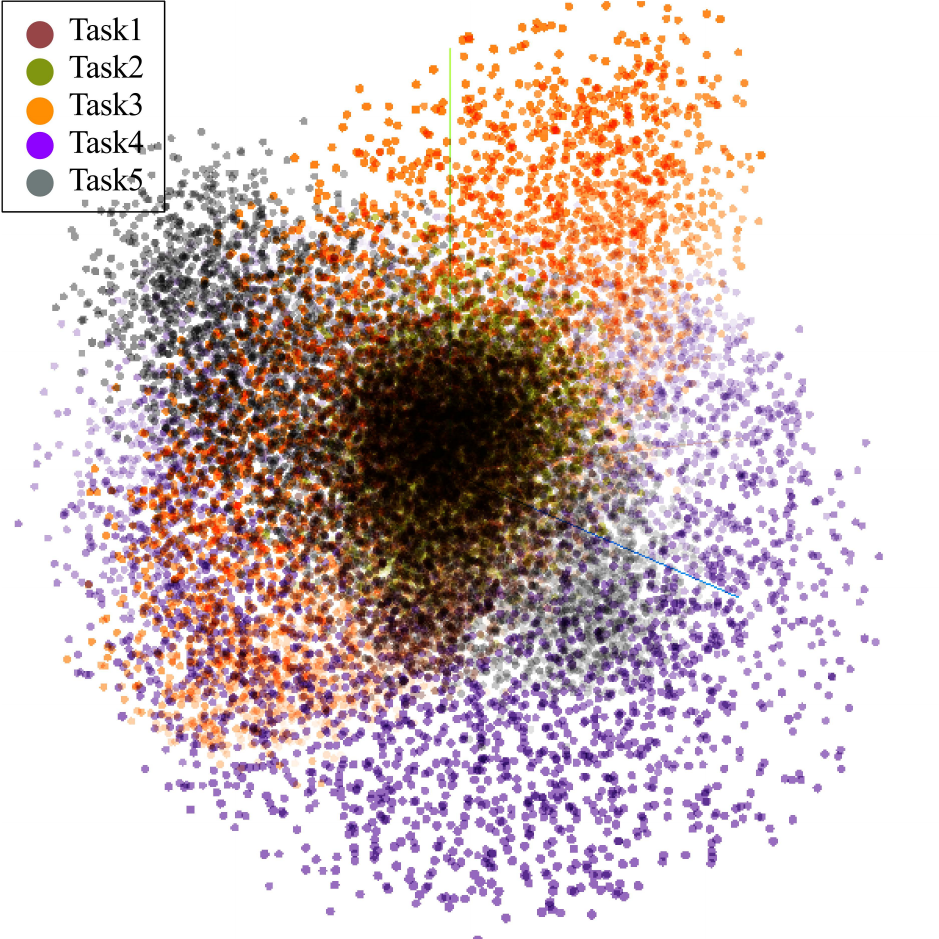}}

    \caption{Visualization of the encoder's output in the S-to-S or S-to-M mode tested with SimpleCNN on the QuaDRiGa-5-subtasks dataset. The potential space is 512-dimensional (CR=1/4), and the sample points for the five subtasks are color-coded.}     
    \label{fig:code-visualization}
\vspace{-0.8cm} 
\end{figure}

Fig. \ref{fig:code-visualization} provides a visualization of the embedding vectors in the S-to-S and S-to-M modes on the QuaDRiGa-5-subtasks dataset. To better observe the distribution characteristics of the data, principal component analysis is applied to map the 512-length vectors into a 2D plane for visualization. Each subtask contains 10,000 samples in the plot. The S-to-M embedding space is looser than that of S-to-S, allowing samples of intersecting subtasks to overlap when mapped to the representation embedding space. The shared encoder creates multiple independent embedding spaces for the compressed code, each corresponding to a task-specific decoder. The S-to-M embedding space can be viewed as five parallel spaces overlapping, with each containing only the corresponding subtask's samples. The shared encoder's ability to represent features is improved by this hard-sharing mechanism, and the multifold expansion of the representation space enables the shared encoder to process and accommodate sample points to be sufficiently dispersed in the representation space. This difference in embedding space explains why points of different colors distribute in different areas in S-to-S but can overlap or be bridged together in S-to-M.

Additionally, the S-to-M mode's superiority over S-to-S can also be explained from the perspective of unsupervised learning in the AE. In S-to-S's unsupervised learning, the visualization in Fig. \ref{5P_simple_mixcode} demonstrates an evident clustering process. The encoder requires significant effort to learn the differences in sample distribution patterns between different subtasks (crossing categories), which results in insufficient attention to the sample differences of the same subtask (within the category). In contrast, S-to-M uses semi-supervised learning, an organic combination of unsupervised and supervised learning, where samples are manually classified by regional labels, eliminating the clustering process of S-to-S. As a result, the encoder can focus more on the sample differences within the category in the regression task.

\section{Conclusion}
\label{conclusion}
We propose a new approach for the deployment mode of CSI feedback DL-based methods called S-to-M mode with GateNet in a MTL-based framework. This mode employs a shared encoder that corresponds to multiple task-specific decoders and a classifier that identifies the task number from the encoder's output. The proposed approach aims to address the issues of UE's computing power or storage capacity consumption, which exist in other modes such as S-to-S or M-to-M. S-to-M mode reduces the NN complexity compared to S-to-S and improves M-to-M by applying MTL, saving UE's memory consumption. The GateNet is designed to address the problem of additional feedback overhead when switching networks, improving the communication system's efficiency and user-friendliness.

In future work, we plan to extend the proposed MTL-based CSI feedback deployment mode to an electronic map, which can be considered as an environmental knowledge base. Each subregion within the global cell can be marked with an electronic map label based on its own environmental scattering characteristic. The S-to-M mode framework can thus provide the electronic map for BS to preload when being first deployed in the cell. Additionally, we plan to develop a more efficient and intelligent environmental electronic map design that considers various factors such as task correlation or resource allocation. We also plan to explore dynamic self-adaptive and self-tuning optimization given the variations of the environment over time. Finally, we aim to extend our deployment mode to the case of implicit CSI feedback in future work.

\bibliographystyle{IEEEtran}
\bibliography{reference}

\begin{thebibliography}{10}
\providecommand{\url}[1]{#1}
\csname url@samestyle\endcsname
\providecommand{\newblock}{\relax}
\providecommand{\bibinfo}[2]{#2}
\providecommand{\BIBentrySTDinterwordspacing}{\spaceskip=0pt\relax}
\providecommand{\BIBentryALTinterwordstretchfactor}{4}
\providecommand{\BIBentryALTinterwordspacing}{\spaceskip=\fontdimen2\font plus
\BIBentryALTinterwordstretchfactor\fontdimen3\font minus
  \fontdimen4\font\relax}
\providecommand{\BIBforeignlanguage}[2]{{%
\expandafter\ifx\csname l@#1\endcsname\relax
\typeout{** WARNING: IEEEtran.bst: No hyphenation pattern has been}%
\typeout{** loaded for the language `#1'. Using the pattern for}%
\typeout{** the default language instead.}%
\else
\language=\csname l@#1\endcsname
\fi
#2}}
\providecommand{\BIBdecl}{\relax}
\BIBdecl

\bibitem{9170653}
F.~{Tariq}, M.~R.~A. {Khandaker}, K.~{Wong}, M.~A. {Imran}, M.~{Bennis}, and
  M.~{Debbah}, ``A speculative study on 6{G},'' \emph{IEEE Wirel. Commun.},
  vol.~27, no.~4, pp. 118--125, Aug. 2020.

\bibitem{5595728}
T.~L. {Marzetta}, ``Noncooperative cellular wireless with unlimited numbers of
  base station antennas,'' \emph{IEEE Trans. Wirel. Commun.}, vol.~9, no.~11,
  pp. 3590--3600, Nov. 2010.

\bibitem{6798744}
L.~Lu, G.~Y. Li, A.~L. Swindlehurst, A.~Ashikhmin, and R.~Zhang, ``An overview
  of massive {MIMO}: Benefits and challenges,'' \emph{IEEE J. Sel. Topics
  Signal Process.}, vol.~8, no.~5, pp. 742--758, Oct. 2014.

\bibitem{4641946}
D.~J. {Love}, R.~W. {Heath}, V.~K. {N. Lau}, D.~{Gesbert}, B.~D. {Rao}, and
  M.~{Andrews}, ``An overview of limited feedback in wireless communication
  systems,'' \emph{IEEE J. Sel. Areas Commun.}, vol.~26, no.~8, pp. 1341--1365,
  Oct. 2008.

\bibitem{kuo2012compressive}
P.-H. Kuo, H.~Kung, and P.-A. Ting, ``Compressive sensing based channel
  feedback protocols for spatially-correlated massive antenna arrays,'' in
  \emph{2012 IEEE Wirel. Commun. Netw. Conf. (WCNC)}.\hskip 1em plus 0.5em
  minus 0.4em\relax IEEE, 2012, pp. 492--497.

\bibitem{2018CsiNet}
C.~{Wen}, W.~{Shih}, and S.~{Jin}, ``Deep {L}earning for {M}assive {MIMO} {CSI}
  {F}eedback,'' \emph{IEEE Wirel. Commun. Lett.}, vol.~7, no.~5, pp. 748--751,
  Oct. 2018.

\bibitem{dong2019deep}
P.~Dong, H.~Zhang, G.~Y. Li, I.~S. Gaspar, and N.~NaderiAlizadeh, ``Deep
  {CNN}-based channel estimation for mm{W}ave massive {MIMO} systems,''
  \emph{IEEE J. Sel. Top. Signal Process.}, vol.~13, no.~5, pp. 989--1000, Jul.
  2019.

\bibitem{ma2020data}
X.~Ma and Z.~Gao, ``Data-driven deep learning to design pilot and channel
  estimator for massive {MIMO},'' \emph{IEEE Trans. Veh. Technol.}, vol.~69,
  no.~5, pp. 5677--5682, Mar. 2020.

\bibitem{sohrabi2020robust}
F.~Sohrabi, H.~V. Cheng, and W.~Yu, ``Robust symbol-level precoding via
  autoencoder-based deep learning,'' in \emph{ICASSP 2020-2020 IEEE
  International Conference on Acoustics, Speech and Signal Processing
  (ICASSP)}.\hskip 1em plus 0.5em minus 0.4em\relax IEEE, Apr. 2020, pp.
  8951--8955.

\bibitem{gao2022data}
Z.~Gao, M.~Wu, C.~Hu, F.~Gao, G.~Wen, D.~Zheng, and J.~Zhang, ``Data-driven
  deep learning based hybrid beamforming for aerial massive {MIMO}-{OFDM}
  systems with implicit {CSI},'' \emph{IEEE J. Sel. Areas Commun.}, vol.~40,
  no.~10, pp. 2894--2913, Aug. 2022.

\bibitem{jiaOverview}
J.~Guo, C.-K. Wen, S.~Jin, and G.~Y. Li, ``Overview of {D}eep {L}earning-based
  {CSI} {F}eedback in {M}assive {MIMO} {S}ystems,'' \emph{IEEE Trans. Commun.},
  vol.~70, no.~12, pp. 8017--8045, Oct. 2022.

\bibitem{zhai2018autoencoder}
J.~Zhai, S.~Zhang, J.~Chen, and Q.~He, ``Autoencoder and its various
  variants,'' in \emph{2018 IEEE Int. Conf. Syst., Man, Cybern. (SMC)}.\hskip
  1em plus 0.5em minus 0.4em\relax IEEE, 2018, pp. 415--419.

\bibitem{lu2020multi}
Z.~Lu, J.~Wang, and J.~Song, ``Multi-resolution {CSI} feedback with deep
  learning in massive {MIMO} system,'' in \emph{Proc. IEEE Int. Conf Commun.
  (ICC)}.\hskip 1em plus 0.5em minus 0.4em\relax IEEE, Jun. 2020, pp. 1--6.

\bibitem{2020GAN}
B.~Tolba, M.~Elsabrouty, M.~G. Abdu-Aguye, H.~Gacanin, and H.~M. Kassem,
  ``Massive {MIMO} {CSI} feedback based on generative adversarial network,''
  \emph{IEEE Commun. Lett.}, vol.~PP, no.~99, pp. 1--1, Dec. 2020.

\bibitem{wang2022transformer}
Y.~Wang, Z.~Gao, D.~Zheng, S.~Chen, D.~Gunduz, and H.~V. Poor,
  ``Transformer-empowered 6{G} intelligent networks: {F}rom massive {MIMO}
  processing to semantic communication,'' \emph{IEEE Wirel. Commun.}, Nov.
  2022.

\bibitem{guo2020convolutional}
J.~Guo, C.-K. Wen, S.~Jin, and G.~Y. Li, ``Convolutional neural network-based
  multiple-rate compressive sensing for massive {MIMO} {CSI} feedback: Design,
  simulation, and analysis,'' \emph{IEEE Trans. Wirel. Commun.}, vol.~19,
  no.~4, pp. 2827--2840, Apr. 2020.

\bibitem{2021EfficientFi-Yang}
J.~Yang, X.~Chen, H.~Zou, D.~Wang, Q.~Xu, and L.~Xie, ``Efficient{F}i: Towards
  large-scale lightweight {W}i{F}i sensing via {CSI} compression,'' \emph{IEEE
  Internet Things J.}, vol.~9, no.~15, pp. 13\,086 -- 13\,095, 01 Aug. 2022.

\bibitem{2018Time-varying}
T.~Wang, C.~K. Wen, S.~Jin, and G.~Y. Li, ``Deep learning-based {CSI} feedback
  approach for time-varying massive {MIMO} channels,'' \emph{IEEE Wirel.
  Commun. Lett.}, vol.~8, no.~2, pp. 416 -- 419, 05 Oct. 2018.

\bibitem{2022Uplink-aided}
J.~Guo, C.-K. Wen, and S.~Jin, ``{CA}net: {U}plink-aided downlink channel
  acquisition in {FDD} massive {MIMO} using deep learning,'' \emph{IEEE Trans.
  Commun.}, vol.~70, no.~1, pp. 199--214, Jan. 2022.

\bibitem{mashhadi2020distributed}
M.~B. Mashhadi, Q.~Yang, and D.~G{\"u}nd{\"u}z, ``Distributed deep
  convolutional compression for massive {MIMO CSI} feedback,'' \emph{IEEE
  Trans. Wirel. Commun.}, vol.~20, no.~4, pp. 2621--2633, Dec. 2020.

\bibitem{sohrabi2021deep}
F.~Sohrabi, K.~M. Attiah, and W.~Yu, ``Deep learning for distributed channel
  feedback and multiuser precoding in {FDD} massive {MIMO},'' \emph{IEEE Trans.
  Wirel. Commun.}, vol.~20, no.~7, pp. 4044--4057, Feb. 2021.

\bibitem{ma2021model}
X.~Ma, Z.~Gao, F.~Gao, and M.~Di~Renzo, ``Model-driven deep learning based
  channel estimation and feedback for millimeter-wave massive hybrid {MIMO}
  systems,'' \emph{IEEE J. Sel. Areas Commun.}, vol.~39, no.~8, pp. 2388--2406,
  Jun. 2021.

\bibitem{COST2100}
J.~Poutanen, K.~Haneda, L.~Liu, C.~Oestges, F.~Tufvesson, and P.~Vainikainen,
  ``Parameterization of the {COST2100} {MIMO} channel model in indoor
  scenarios,'' in \emph{Proc. Eur. Conf. Antennas Propag.}, Apr. 2011, pp.
  3606--3610.

\bibitem{xiao2021ai}
H.~Xiao, Z.~Wang, W.~Tian, X.~Liu, W.~Liu, S.~Jin, J.~Shen, Z.~Zhang, and
  N.~Yang, ``{AI} enlightens wireless communication: Analyses, solutions and
  opportunities on {CSI} feedback,'' \emph{China Commun.}, vol.~18, no.~11, pp.
  104--116, Nov. 2021.

\bibitem{QuaDRiGa}
S.~{Jaeckel}, L.~{Raschkowski}, K.~{Börner}, and L.~{Thiele},
  ``{Q}ua{DR}i{G}a: A 3-{D} multi-cell channel model with time evolution for
  enabling virtual field trials,'' \emph{IEEE Trans. Antennas Propag.},
  vol.~62, no.~6, pp. 3242--3256, Jun. 2014.

\bibitem{TransferCSI2021Zeng}
J.~Zeng, J.~Sun, G.~Gui, B.~Adebisi, T.~Ohtsuki, H.~Gacanin, and H.~Sari,
  ``Downlink {CSI} feedback algorithm with deep transfer learning for {FDD}
  massive {MIMO} systems,'' \emph{IEEE Trans. Cogn. Commun. Netw.}, vol.~7,
  no.~4, pp. 1253--1265, Dec. 2021.

\bibitem{Zhang2023MTL}
B.~Zhang, H.~Li, X.~Liang, X.~Gu, and L.~Zhang, ``Multi-task training approach
  for {CSI} feedback in massive {MIMO} systems,'' \emph{IEEE Commun. Lett.},
  vol.~27, no.~1, pp. 200--204, 2023.

\bibitem{2021Beamforming}
R.~Hu, L.~Jiang, and P.~Li, ``Hybrid beamforming with deep learning for
  large-scale antenna arrays,'' \emph{IEEE Access}, vol.~9, pp.
  54\,690--54\,699, Mar. 2021.

\bibitem{2020Deep-Transfer}
Y.~Yang, F.~Gao, Z.~Zhong, B.~Ai, and A.~Alkhateeb, ``Deep transfer
  learning-based downlink channel prediction for {FDD} massive {MIMO}
  systems,'' \emph{IEEE Trans. Commun.}, vol.~68, no.~12, pp. 7485--7497, Aug.
  2020.

\bibitem{2018Channel-Estimation}
H.~Xie, F.~Gao, S.~Jin, J.~Fang, and Y.-C. Liang, ``Channel estimation for
  {TDD}/{FDD} massive {MIMO} systems with channel covariance computing,''
  \emph{IEEE Trans. Wirel. Commun.}, vol.~17, no.~6, pp. 4206--4218, Jun. 2018.

\bibitem{2000system-model}
K.~Pedersen, P.~Mogensen, and B.~Fleury, ``A stochastic model of the temporal
  and azimuthal dispersion seen at the base station in outdoor propagation
  environments,'' \emph{IEEE Trans. Veh. Technol.}, vol.~49, no.~2, pp.
  437--447, Mar. 2000.

\bibitem{2013Spatially}
O.~E. Ayach, S.~Rajagopal, S.~Abu-Surra, Z.~Pi, and R.~Jr, ``Spatially sparse
  precoding in millimeter wave {MIMO} systems,'' \emph{IEEE Trans. Wirel.
  Commun.}, vol.~13, no.~3, pp. 1499--1513, Mar. 2013.

\bibitem{2017Channel}
A.~Alkhateeb, O.~E. Ayach, G.~Leus, and R.~W. Heath, ``Channel estimation and
  hybrid precoding for millimeter wave cellular systems,'' \emph{IEEE J. Sel.
  Topics Signal Process.}, vol.~8, no.~5, pp. 831--846, Oct. 2017.

\bibitem{ruder2017overview}
S.~Ruder, ``An overview of multi-task learning in deep neural networks,''
  \emph{arXiv preprint arXiv:1706.05098}, 2017.

\bibitem{collobert2008Hardsharing}
R.~Collobert and J.~Weston, ``A unified architecture for natural language
  processing: Deep neural networks with multitask learning,'' in \emph{Proc.
  25th Int. Conf. Mach. Learn.}, Jul. 2008, pp. 160--167.

\bibitem{misra2016cross}
I.~Misra, A.~Shrivastava, A.~Gupta, and M.~Hebert, ``Cross-stitch networks for
  multi-task learning,'' in \emph{Proc IEEE Comput Soc Conf Comput Vision
  Pattern Recognit.}, 2016, pp. 3994--4003.

\bibitem{baxter1997bayesian}
J.~Baxter, ``A bayesian/information theoretic model of learning to learn via
  multiple task sampling,'' \emph{Mach. Learn.}, vol.~28, no.~1, pp. 7--39,
  Jul. 1997.

\end{thebibliography}
\end{document}